\documentclass[acmlarge,table,screen]{acmart}
\renewcommand\footnotetextcopyrightpermission[1]{}

\pdftrailerid{}
\usepackage[T1]{fontenc}

\usepackage{amsmath}
\usepackage{wasysym}
\usepackage{fdsymbol}
\usepackage{xcolor}
\usepackage{array,multirow,multicol,ragged2e}
\usepackage{tabularx}
\usepackage{threeparttable}
\usepackage{makecell}
\usepackage{circledsteps}
\usepackage{graphicx}
\graphicspath{{img/}}

\usepackage{rotating}
\usepackage{diagbox}

\usepackage{enumitem}
\usepackage{tikz}
\usepackage{tikz-cd}
\usepackage[all]{xy}
\usetikzlibrary{tikzmark,shapes,arrows}

\tikzstyle{decision} = [diamond, draw, 
text width=4.5em, text badly centered, node distance=3cm, inner sep=0pt]
\tikzstyle{block} = [rectangle, draw, align=center, text width=14em]
\tikzstyle{line} = [draw, -latex']
\tikzstyle{cloud} = [draw, ellipse,fill=red!20, node distance=3cm,
minimum height=2em]

\newcommand{\mcrot}[4]{\multicolumn{#1}{#2}{\rlap{\rotatebox{#3}{#4}~}}} 

\newcommand{\subhead}[1]{\vspace{2pt}\noindent{\textbf{#1.}}}
\newcommand{\subheadit}[1]{\vspace{4pt}\noindent{\textit{#1.}}}

\newcommand{\yes}{{\small\CIRCLE}}
\newcommand{\no}{{\small\Circle}}
\newcommand{\hmm}{{\small\LEFTcircle}}
\newcommand{\maybe}{{\small\RIGHTcircle}}

\newcommand{\killedByI}[1]{\textcolor{lightgray}{#1}}

\renewcommand*{\textsf}[1]{{\small\sffamily#1}}

\newcolumntype{L}[1]{>{\raggedright\let\newline\\\arraybackslash\hspace{0pt}}m{#1}}
\newcolumntype{C}[1]{>{\centering\let\newline\\\arraybackslash\hspace{0pt}}m{#1}}
\newcolumntype{R}[1]{>{\raggedleft\let\newline\\\arraybackslash\hspace{0pt}}m{#1}}

\newcommand{\TLSMsg}[1]{\Circled[fill color=gray!30,inner color=black]{{\normalfont\footnotesize\sffamily#1}}}

\setcopyright{none}

\acmJournal{CSUR}
\acmVolume{}
\acmNumber{}
\acmArticle{}
\acmMonth{0}

\begin{document}

\author{Xavier \lowercase{de} Carné \lowercase{de} Carnavalet}
\email{xavier.decarnedecarnavalet@polyu.edu.hk}
\affiliation{%
 \institution{The Hong Kong Polytechnic University} 
 \streetaddress{11 Yuk Choi Rd}
 \city{Hong Kong}
 \state{Kowloon}
 \postcode{Hung Hom}
 \country{Hong Kong SAR}
}
\author{Paul C.\ \lowercase{van} Oorschot}
\email{paulv@scs.carleton.ca}
\affiliation{%
 \institution{Carleton University, Ontario} 
 \streetaddress{1125 Colonel By Drive, Ottawa}
 \city{Ottawa}
 \state{ON}
 \postcode{K1S 5B6}
 \country{Canada}
}
\authorsaddresses{Authors' address: Xavier de Carné de Carnavalet, Department of Computing, The Hong Kong Polytechnic University, 11 Yuk Choi Rd, Hung Hom, Kowloon, Hong Kong; email: xavier.decarnedecarnavalet@polyu.edu.hk; Paul C. van Oorschot, School of Computer Science, Carleton University, 1125 Colonel By Drive, Ottawa, ON, K1S 5B6, Canada; email: paulv@scs.carleton.ca. Most of this work was conducted when the first author was a postdoctoral researcher at Carleton University.}

\renewcommand{\shortauthors}{Xavier de Carné de Carnavalet and Paul C. van Oorschot}

\title[A survey on TLS interception]{A survey and analysis of TLS interception\\ mechanisms and motivations}
\titlenote{A version of this paper will appear in: \textit{ACM Computing Surveys}.}
\subtitle{Exploring how end-to-end TLS is made ``end-to-me'' for web traffic}

\begin{abstract}
TLS is an end-to-end protocol designed to provide confidentiality and integrity guarantees that improve end-user security and privacy. While TLS helps defend against pervasive surveillance of intercepted unencrypted traffic, it also hinders several common beneficial operations typically performed by middleboxes on the network traffic. Consequently, various methods have been proposed that ``bypass'' the confidentiality goals of TLS by playing with keys and certificates essentially in a man-in-the-middle solution, as well as new proposals that extend the protocol to accommodate third parties, delegation schemes to trusted middleboxes, and fine-grained control and verification mechanisms. We first review the use cases expecting plain HTTP traffic and discuss the extent to which TLS hinders these operations. We retain 19 scenarios where access to unencrypted traffic is still relevant and evaluate the incentives of the stakeholders involved. Second, we survey 30 schemes by which TLS no longer delivers end-to-end security, and by which the notion of an ``end'' changes, including caching middleboxes such as Content Delivery Networks. Finally, we compare each scheme based on deployability and security characteristics, and evaluate their compatibility with the stakeholders' incentives. Our analysis leads to a number of key findings, observations, and research questions that we believe will be of interest to practitioners, policy makers and researchers.
\end{abstract}


\maketitle
\thispagestyle{empty}
\fancyfoot{}

\section{Introduction}
End-user security and privacy have improved due to the recent sharp increase in end-to-end TLS-encrypted web traffic~\cite{KotziasRAPVC18}, i.e., HTTPS traffic (Hypertext Transfer Protocol (HTTP) with the Secure Socket Layer (SSL)/Transport Layer Security (TLS)).
However, as the Internet traffic originally carried mostly plaintext communications, many network-related practices have been built relying on the fact that the plaintext data is readily accessible~\cite{rfc8404}.
Unsurprisingly, and contrary to some stakeholders view, the shift towards HTTPS is reported to disturb a number of legitimate operations regularly performed by software or hardware middleboxes on plaintext network traffic, including~\cite{draft-camwinget-tls-use-cases-05,rfc8404,bits-ets}: network management and monitoring, problem troubleshooting, performance optimization, caching, intrusion detection, malware download and data leakage prevention, fraud monitoring, parental controls.

As a first solution that comes to mind, many client-facing security products and enterprise network gateways intercept HTTPS traffic by terminating and re-creating TLS sessions to access the plaintext~\cite{xcc-tls-proxy,https-intercept-impact}.
However, unlike with plain HTTP, any means of intercepting HTTPS traffic is seen as a threat by many stakeholders, on top of being precisely what TLS is meant to prevent.
This conflict was illustrated during the standardization of TLS 1.3, whereby some industry participants pushed for means to facilitate traffic inspection, while privacy and security advocates raised concerns about numerous weaknesses introduced by middleboxes in the past~\cite{symantec-tls-responsible,tinfoil}. In the end, TLS 1.3 was made less friendly to passive monitoring (by removing non-forward secret ciphersuites), resulting e.g., in the banking industry to promote as a competing standard an interception-friendly protocol: Enterprise TLS (ETS), opposed by, e.g., the Electronic Frontier Foundation~\cite{ets-eff}.

Industry actors have sought to minimize the deployment cost of traffic interception by focusing on basic technical solutions compatible with TLS, e.g., sharing decryption keys (e.g., by CDNs below), trusting custom root certificates. Those solutions provide full access to the plaintext and little oversight, which severely puts at risk the benefits of TLS.
Thus, even when a middlebox is allowed to inspect HTTPS traffic, it is desirable to limit/audit its privileges by providing restricted access (e.g., read but not write, with the URL path visible but not the full request or body, for a limited time), or attesting the middlebox software. 
To this end, various groups have proposed or developed TLS variants to accommodate third parties.

Moreover, the emergence of Content Distribution Networks (CDNs) extends the notion of traffic interception and poses significant challenges to security and privacy.
In particular, for CDNs to serve content over HTTPS, it is common for the content provider to either share their certificate private key with the CDN, or allow the CDN to obtain certificates for their domain (e.g., by pointing the domain name to the CDN itself)~\cite{https-meets-cdn}.
Therefore, CDNs change the notion of the ``end'' in end-to-\emph{end} encrypted TLS sessions, pushing for new ways to represent an authorization or delegation to the CDN as a (un)trusted third party.

\subhead{Contributions}
In this paper, before we jump into a technical analysis of proposals from the literature, we first lay down the necessary foundations by reviewing common network practices that rely on access to plaintext HTTP traffic, and the documented impact of TLS on them.
We notably rely on RFC 8404~\cite{rfc8404} to document the industry's perspective on the \emph{pervasive encryption} of data at-rest and in-transit, and distill meaningful arguments with respect to TLS.
We discuss the extent of the impact of TLS, especially TLS 1.3, and identify 19 actual use cases that benefit or require access to the unencrypted traffic, which we sort into four categories based on their underlying motivation: legal, security/privacy, performance, and others.
We also evaluate the incentives of the six main stakeholders that support these use cases (e.g., end user, ISP, content provider). Importantly, no use cases are simultaneously supported by both client-end and server-end stakeholders, although the participation of both is often needed.

Then, we examine the techniques and proposals that aim at introducing a third party (``me'') in the communication for traffic inspection or enabling passive monitoring, i.e., \emph{end-to-me} communications.
Through a thorough literature review and our familiarity with the literature, we select for discussion the most prominent schemes, and also consider the 30 most viable ones and those with the most interesting features.
We thus review various ways by which long-term keys, session keys or intermediate secrets are shared, especially in the context of CDNs. We also explore the ways to achieve delegation with special-purpose certificates or tokens, and methods proposed to guarantee the secrecy of middlebox operations.
Variants of TLS that handshake with middleboxes to achieve fine-grained control are also discussed. We use the term TLS ``interception'' broadly to cover all these schemes.
We propose the following taxonomy for all schemes, each mapped to sections in this paper: session splitting and key sharing (§\ref{sec:enterprise-twists}), content delegation (§\ref{sec:cdn}), three-way handshakes (§\ref{sec:universal-protocols}), and privacy-preserving inspection (§\ref{sec:privacy-inspection}).

Furthermore, we evaluate the scheme compatibility with the stakeholders' incentives. Surprisingly, many proposals are inapplicable due to obvious incompatible incentives of the relevant parties.
Finally, we provide a comparative evaluation of these schemes based on 27 security and deployment criteria.
As an example of our findings, our analysis leads to the conclusion that a simpler design addressing multiple problems at once is more likely to be deployed, and that there is a lack of a secure and robust middlebox discovery feature.

The rest of the paper is organized as follows. We first describe regular network operations that expect access to plaintext HTTP (§\ref{sec:http-access}), and discuss the extent of the impact of TLS on them (§\ref{sec:challenges-tls}).
We provide a stakeholder incentives analysis in §\ref{sec:stakeholders}.
In §\ref{sec:enterprise-twists}, we explore the most common techniques employed by enterprises and client software to gain access to the plaintext of HTTPS traffic, along with prominent new proposals such as ETS.
Caching middleboxes constitute a unique problem, for which we discuss related delegation proposals in §\ref{sec:cdn}.
Proposals that deeply change TLS to make it friendly to middleboxes are discussed in §\ref{sec:universal-protocols}.
Ways to reduce trust requirements on middleboxes are summarized in §\ref{sec:privacy-inspection}.
Our systematization leads to the categorization and evaluation of the techniques and proposals against a 27-criteria framework in §\ref{sec:evaluation}, with various insights in §\ref{sec:discussion}.
Among our main contributions are summaries provided in Table~\ref{tab:proposal-usage} and Table~\ref{tab:proposal-eval-cdn}. The rest of the paper provides background and information to support and understand our analysis.

\section{SSL/TLS: Background}
\label{sec:background}
The reader familiar with TLS may choose to skip this section.
We provide herein a review of important aspects of the TLS protocol (formerly SSL), which are helpful to understand the rest of the paper. We focus on the latest major revisions only: versions 1.2 and 1.3, and we do not discuss exotic configurations. The interested reader is referred to related RFCs~\cite{rfc6101,rfc2246,rfc4346,rfc5246,rfc8446}, books and publications~\cite{rescola-ssl,clark-sok,ristic2014bulletproof}, \cite[Chapter 6]{BoydMS20} for more details.

\subsection{TLS handshake overview}
The TLS protocol performs an authenticated key exchange (or key transport) during an initial handshake to establish a shared secret from which symmetric encryption keys are derived. Prior to TLS 1.3, this key exchange is either based on RSA, Finite Field Diffie-Hellman (DH) or Elliptic Curve DH (ECDH). TLS 1.3 removed the RSA option. (EC)DH only uses ephemeral keys in 1.3 or may do so in prior versions, and is then referred as (EC)DHE. Ephemeral key exchanges provide (perfect) forward secrecy or PFS, i.e., an attacker with the knowledge of the server's long-term key cannot decrypt past recorded sessions as the ephemeral private key has been deleted.

In the RSA key transport, a pre-master secret (PMS) is chosen by the client and encrypted using the server's RSA long-term public key. The server then decrypts the encrypted PMS (ePMS) and derives the shared master secret (MS) from the PMS and client/server-provided random nonces. With DH key exchanges, the server sends a public key share signed with its long-term private key. The client verifies the signature and combines the server's key share with its own private key share to derive the PMS. Conversely, the server combines its private key share with the client's public key share. Once the handshake is finished, the rest of the communication is encrypted with keys derived from the MS.

\subsection{Protocol messages}
\label{sec:background-messages}
We describe below the messages exchanged during the TLS handshake to achieve key exchange/transport.

\subsubsection{\textsf{ClientHello} \TLSMsg{CH}}
The client begins the handshake by sending a \textsf{ClientHello}.
Depending on the TLS version, the format of these messages slightly differs.

\subhead{TLS 1.2} The \textsf{ClientHello} lists the maximum TLS version supported, a random number (\textsf{client\_random}), a session ID when resuming a previous session, and a list of ciphersuites ordered by preference that specify cryptographic primitives that the client supports for key exchange, encryption, and message authentication.
As mentioned above, the key exchange is either RSA, or (EC)DH(E). The encryption ciphers could be e.g., AES or ChaCha20. The mode of operation for block ciphers is also specified, and either provides authenticated encryption with additional data (AEAD, e.g., GCM and CCM) or not (e.g., CBC). Finally, the message authentication is performed using an HMAC based on e.g., SHA256 or SHA384.

The \textsf{ClientHello} also includes a list of TLS extensions. Notable ones include the \textsf{server\_name\_indication} (SNI) that specifies the server name (i.e., domain name); \textsf{supported\_groups} with a list of supported elliptic curves for the (EC)DHE key exchange, e.g., SECP256r1, SECP384r1, x25519; and \textsf{signature\_algorithms} with a list of supported signature algorithms used to authenticate the (EC)DHE parameters, e.g., RSA or ECDSA, with various padding and message digest algorithms.

\subhead{TLS 1.3} The \textsf{ClientHello}'s maximum supported TLS version field is set to 1.2 while all the supported versions are indicated in a separate TLS extension to avoid middlebox intolerance to the newer protocol. Similarly, the session ID is set to a random value to mimic a session resumption, while the feature is achieved through a separate Pre-Shared Key (PSK) TLS extension.
The ciphersuites are different than in TLS 1.2 as each algorithm is selected separately. The ciphersuite no longer specifies the primitives for key exchange, which is forced to (EC)DHE. All the ciphersuites include AEAD ciphers.
The \textsf{ClientHello} in TLS 1.3 also includes the client's key share based on its preferred group/elliptic curve.

\subsubsection{\textsf{ServerHello} \TLSMsg{SH}}
The server selects the highest version it supports and that the client also advertised. It choses among the client's ciphersuites (respecting or not the client's preferred order), ECs, and signature algorithms. In TLS 1.3, if the server does not support the client's chosen EC, it sends a \textsf{HelloRetryRequest} with a list of the server's supported curves, and the client resends a \textsf{ClientHello} with a different key share.

The server then answers the client with a \textsf{ServerHello} that indicates the chosen parameters, a random value (\textsf{server\_random}), and an optional session ID for the client to resume the session at a later time prior to TLS 1.3. The server also includes its key share in TLS 1.3, at which point both the client and the server can already derive a shared secret, called \textsf{handshake\_secret}. Note that this key exchange is not yet authenticated. Further extensions are sent encrypted in TLS 1.3 in an \textsf{EncryptedExtensions} \TLSMsg{EE} message.

\subsubsection{\textsf{Certificate} \TLSMsg{CRT}}
In TLS 1.2, the server sends a \textsf{Certificate} message after the \textsf{ServerHello} with the server's certificate chain. The chain is composed at least of an end-entity certificate (EEC, also called a leaf certificate). Additional intermediate Certificate Authority (CA) certificates may be included for clients to establish a chain to a trusted root CA.
In TLS 1.3, the Certificate and subsequent handshake messages are encrypted using keys derived from the \textsf{handshake\_secret}. The (EC)DHE parameters are authenticated next by the \textsf{CertificateVerify}~\TLSMsg{CV} message that contains a hash of the handshake messages (transcript $\tau$) signed by the server's certificate private key.
In TLS 1.3, each certificate can be accompanied by certificate extensions, e.g., OCSP status (see §\ref{sec:cert-validation}).

\subsubsection{\textsf{KeyExchange}}
Prior to TLS 1.3, when a non-RSA ciphersuite is negotiated, the server then sends its signed key share in the \textsf{ServerKeyExchange} \TLSMsg{SKE} message and ends with an empty \textsf{ServerHelloDone} \TLSMsg{SHD} message. The client responds with its key share in a \textsf{ClientKeyExchange} \TLSMsg{CKE} message, wasting round trips compared to TLS 1.3.
For RSA ciphersuites, the \textsf{ServerKeyExchange} is skipped, and the client simply sends the ePMS in the \TLSMsg{CKE} message.

\subsubsection{\textsf{ChangeCipherSpec} \TLSMsg{CCS}}
When one end has established the shared encryption keys and plans to send the following messages encrypted, it sends a notification in the form of a \textsf{ChangeCipherSpec} message. This message has been made optional in TLS 1.3.

\subsubsection{\textsf{Finished} \TLSMsg{FIN}}
Finally, both the server and then the client send to each other a MAC (\texttt{verify\_data}) of all the handshake messages exchanged ($\tau$), encrypted with the appropriate derived keys as described below in §\ref{sec:channel-enc-keys}.

\subsection{Channel encryption keys}
\label{sec:channel-enc-keys}
Once a MS has been established, the two ends derive a number of keys. The client and server encrypt their messages with a separate set of symmetric keys, MAC keys and Initialization Vectors (IVs), respectively called \textsf{client\_write\_\{key,MAC\_key,IV\}} and \textsf{server\_write\_\{key,MAC\_key,IV\}}.
When using AEAD ciphers, the \textsf{\{client,server\}\_write\_key} serve as the MAC keys.

The derivation of these keys from the MS is based on a Pseudorandom Function  (PRF) prior to TLS 1.3 as defined in the TLS standard, while version 1.3 leverages another construction based on the HKDF defined in RFC5869~\cite{rfc5869}.
In TLS 1.3, a shared secret, \textsf{handshake\_secret}, is already established early during the handshake. Four keys and IVs are derived to encrypt the rest of the handshake messages, i.e., \textsf{\{client,server\}\_handshake\_\{key,iv\}}, in a similar fashion to TLS 1.2 encryption keys.
The MS is also derived from \textsf{handshake\_secret}, and leads to separate application keys and IVs, i.e., \textsf{\{client,server\}\_application\_\{key,iv\}}, which in turn encrypt the rest of the communication.

\subsection{Certificate validation}
\label{sec:cert-validation}
When validating the server certificate, the client tries to chain the certificates it received to a root CA certificate it already trusts, by verifying the signatures. The list of trusted CAs is typically preloaded with the operating system or individual browsers and updated periodically.
Then, the client checks whether the leaf certificate is appropriate for the contacted domain for web purposes, and that all certificates in the chain are not expired or revoked (through e.g., CRL, OCSP).
Certain browsers also mandate that the certificates be recorded in multiple Certificate Transparency (CT) logs. This is verified by the presence of a Precertificate Signed Certificate Timestamp (SCT) issued by the logs. The SCT is either embedded in the final certificate or provided to the client through a TLS extension.
Finally, the client may require that only specific leaf or issuing certificates be used, by pinning their keys (cf.\  HPKP~\cite{rfc7469}).

\section{Operations expecting HTTP access}
\label{sec:http-access}
\begin{figure}[t]
\includegraphics[width=.8\columnwidth]{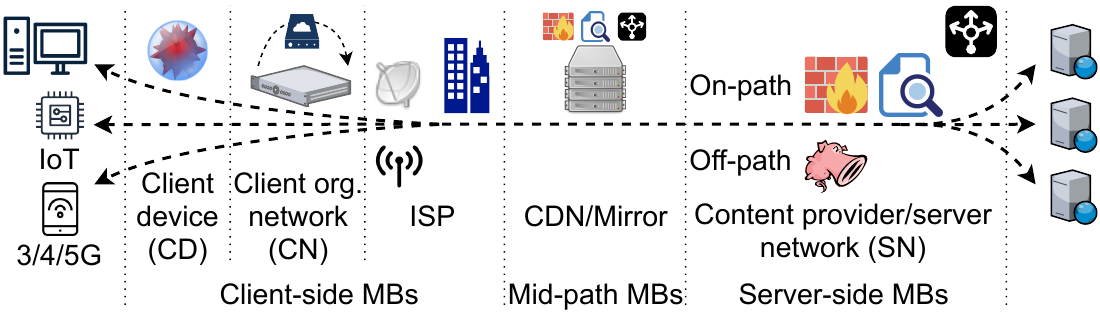}
\caption{Location of middleboxes (MB) impacted by TLS on the connection path}
\label{fig:middleboxes}
\end{figure}

This section builds the foundations for the rest of the paper by listing the common network practices that rely on access to plain HTTP traffic. These practices are positioned as the reason that TLS interception is needed. 
We identify four main categories of use cases: legal, security, performance-related and other reasons.
Use cases are negatively impacted or precluded by the deployment of HTTPS. 
An open question for consideration is: do these practices (use cases) provide sufficient justification for HTTPS interception, or is interception simply the easiest way to accommodate the existing practices?
This actual impact of TLS on these practices is discussed in §\ref{sec:challenges-tls}.
We group middleboxes operating on the traffic based on their location on the network path between the endpoints, i.e., client-side (up to the user's ISP), mid-path (including CDNs, although edge servers are usually located closer to the user), and server-side. Figure~\ref{fig:middleboxes} illustrates the location of these middleboxes on the network.

\subsection{Government-driven ``legal'' common practices for HTTP access}
We describe below legal requirements for processing cleartext network traffic as it pertains to web traffic.

\subheadit{Lawful interception}
Lawful interception is a constraint that a government can impose on operators to obtain a specific portion of the traffic they are carrying, typically upon delivering a warrant or subpoena. Specific architectures and protocols for lawfully intercepting IP traffic have been standardized~\cite{rfc3924,li-etsi}.
Note that this practice is distinct from warrantless surveillance (cf.\ Snowden's revelations~\cite{nsa-guardian}), which we do not consider here as a use case.

\subheadit{ISP data retention}
Internet service providers (ISP) may be legally required to keep network traffic records to comply with data retention laws. Those laws vary per country. For instance, in the European Union (EU), the Data Retention Directive was adopted in 2006 and specified the retention from six months to two years of various pieces of metadata. This directive was invalidated in 2014 by the Court of Justice of the EU~\cite{cjeu-data-retention}; however, numerous EU countries continue to apply their own retention laws~\cite{stopretention}. In Australia, an equivalent data retention law has passed in 2015~\cite{australia-isp-law}. In the United States, there is no law to our knowledge that mandates ISPs to retain data.

\subheadit{Content filtering mandated by governments}
A government may impose content filtering policies on network operators~\cite{rfc8404}. The policies could be targeted at blocking access to illegal websites, related to e.g., child abuse, content piracy.\footnote{See the blocking of The Pirate Bay. \url{https://en.wikipedia.org/wiki/Countries_blocking_access_to_The_Pirate_Bay}}
Content filtering may be implemented at various layers~\cite{is-content-blocking}: 1)~blocking access to certain IP addresses or to certain protocols; 2)~blocking DNS resolution to targeted domains; 3)~recognizing content to be blocked by keyword matching or fingerprinting from a blacklist; 4) blocking known URLs; and 5) delisting the websites from search engines. Only \#3 and \#4 assume access to cleartext web traffic.

\subheadit{Fraud detection}
Certain industries perform fraud monitoring (e.g., stock fraud) of certain employees as part of out-of-band traffic inspection~\cite{draft-rhrd-tls-tls13-visibility-01,bits-ets}.

\subsection{Security-based common practices for HTTP access}
We list below some prominent examples of security-related practices that also rely on cleartext traffic.

\subheadit{Malware/intrusion detection/prevention}
Network security functions, e.g., Intrusion Detection/Prevention Systems (IDS/IPS) and Web Application Firewalls (WAF), control network communications in search of attack patterns or to detect vulnerabilities~\cite{draft-camwinget-tls-use-cases-05,bits-ets}. They require access to the application-layer traffic.
Also, antivirus (AV) solutions control the traffic to prevent malware downloads and to block known malicious URLs~\cite{bits-ets,draft-camwinget-tls-use-cases-05}.

\subheadit{Data retention for compliance}
Organizations and data center operators may also want to keep network packet captures to investigate attacks (forensics), or demonstrate compliance to industry standards~\cite{rfc8404}. Note that such industry standards may not be government-driven and are not enforced by law.

\subheadit{DDoS mitigation}
CDNs may also provide, beyond simply caching content, volumetric Distributed Denial-of-Service (DDoS) attack protection. To enable access to users who were incorrectly flagged as attackers, CDNs sometimes serve a CAPTCHA challenge before the request is honored to differentiate between human and automated queries (also called ``automatic bot discernment''). This mechanism is also used to block anonymous queries to reach a website, e.g., when using Tor (see~\cite{KhattakFAJSMPM16}). 
Application-layer (L7) DDoS attacks are more difficult to automatically detect than volumetric DDoS attacks. The former requires inspection of the application-layer traffic to understand that e.g., a resource-hungry feature of a website is being abnormally triggered.

\subheadit{Data loss prevention}
Companies may want to monitor outgoing traffic to stop data exfiltration and prevent information loss~\cite{transitive-trust-dell,draft-camwinget-tls-use-cases-05,bits-ets}.
These tools try to identify (possibly obfuscated) sensitive data in the traffic.

\subsection{Performance-based common practices for HTTP access}
Several optimization mechanisms rely on information about the traffic to improve performance-related decisions, ranging from load balancing and Quality of Service (QoS) adaptation, to data broadcasting and caching.

\subheadit{Application load balancing}
Network loads can be balanced in a number of ways to back-end servers. A fine-grained approach consists in making decisions at the application layer, e.g., based on the request path (cf.\ AWS Application Load Balancer (ALB)~\cite{esc-lb-types}).
According to RFC8404~\cite{rfc8404}, cellular networks may route traffic depending on application-layer criteria. This could be the case e.g., as part of the Mobility Load Balancing (MLB) function of 3GPP Self-Organizing Networks (SONs) where congested cells can hand over some load to other less busy cells.
Another cellular network optimization consists in queuing traffic based on its bandwidth and latency needs. To accurately detect the type of application traffic, signatures are matched by a monitoring system that performs deep-packet inspection (DPI).
Finally, for video streaming media, a dynamic QoS can be adapted based on the video bit-rate information obtained from plaintext HTTP requests (e.g., as done in~\cite{adaptive-qos}).

\subheadit{Broadcasting}
Content broadcasting helps reduce bandwidth requirements by sending the data once for a given region, which is then locally redistributed to all clients.
This is especially useful in satellite communications where a satellite can broadcast the signal to many receivers~\cite{satellite-https}.
Similarly, the 3GPP Multimedia Broadcast/Multicast Services (MBMS) specification aims at distributing live TV content to trusted edge proxies, to be delivered to the final users through either unicast or multicast.

\subheadit{Caching and replication}
Caching popular content near the network edge is a way to reduce unnecessary bandwidth consumption and latency. Content caching and replication may be performed upon agreement between the content service provider and CDNs, whereby the former delegates its content to CDNs. 
Caching may also be transparent to the end parties, when performed e.g., by ISPs~\cite{mctls,rfc8404}, although this practice is controversial~\cite{rfc8404}. 

\subheadit{Transcoding and compression}
Mobile devices with slow connections, limited data caps, short battery life or slow CPUs may benefit from transparent proxies that (re)compress web contents in an effort to minimize transfer and processing times~\cite{rfc8404}. This service can be built into browsers, e.g., Opera Mini~\cite{opera-mini}, Google's Flywheel Chrome extension~\cite{flywheel}. Such performance-enhancing proxies can downsize image resolution, video and audio quality~\cite{mctls,web-proxies}. Mobile network operators have been shown to compress images~\cite{imc-internet-violations}.
Also, performance-enhancing proxies at the network edge, as used on mobile or satellite links, could ``further control pacing, limit simultaneous HD videos, or prioritize active videos against new videos''~\cite{rfc8404}.

\subsection{Other common practices motivating HTTP access}
Here we list other cases that do not fit into the previous three categories.

\subheadit{Diagnosing and troubleshooting issues} 
Network packet captures traditionally provide visibility into networks, assist in troubleshooting problems, and are used to analyze application performance~\cite{rfc8404,draft-green-tls-static-dh-in-tls13-01}. Captures can even be provided by customers as taken close to the client-side, allowing for a targeted and efficient troubleshooting.
Observing actual network communications is often the easiest, when not the last, resort to troubleshoot a problem.

\subheadit{Content filtering}
Parental control solutions also filter network traffic to block forbidden contents, remove explicit images, sensor swear words~\cite{xcc-tls-proxy}. Similarly, ad blockers may choose to remove unwanted ads from the web traffic before it reaches the browser.
Finally, employers may want to regulate the use of their resources by employees by, e.g., collecting browsing histories, preventing abusive use of bandwidth~\cite{draft-camwinget-tls-use-cases-05}.
Similar filtering needs to legal constraints on ISPs may apply.

\subheadit{Content-based billing}
Network operators may leverage fine-grained billing of their customers based on the website or URL visited to, e.g., provide free-of-charge access to certain contents (zero-rating~\cite{choffnes2017empirical}), or charge more for premium services~\cite{cisco-ecs}.

\subheadit{Other/abusive practices}
ISPs may insert tracking headers~\cite{header-enrichement,verizon-supercookie}, with or without personally identifiable information such as a phone's IMEI and IMSI. Sometimes, this is simply intended to pass information between internal parts of the operator's network; however, this may also be done for advertisement and tracking purposes~\cite{header-enrichement,rfc8404,cisco-ecs}. ISPs could also insert interstitials to inform users of e.g., quota limits~\cite{rfc8404}.
Finally, they may also monetize HTTP errors by inserting ads on error pages~\cite{web-proxies}.
We do not further consider these controversial practices by ISPs in this study due to the absence of benefits to the end user (without other easy alternatives), and lack of any legal requirement.
~\\[.5em]
\noindent The impact of the above-mentioned cases is discussed next.

\section{Challenges introduced by TLS}
\label{sec:challenges-tls}

We discuss in this section the impact of TLS on the aforementioned use cases (§\ref{sec:http-access}), as reported in IETF and industry discussions~\cite{draft-camwinget-tls-use-cases-05,rfc8404,bits-ets,draft-rhrd-tls-tls13-visibility-01}, and academic papers~\cite{NaylorFLGMMPS14,mctls}.
We hint at obvious alternative solutions when such use cases can be achieved without impeding TLS.
We also observed that regulations are often cited throughout these references as a major use case. We address this particular concern separately.
Finally, we identify 19 use cases for access to HTTP traffic in the presence of TLS.

\subsection{Evaluation of HTTP access use cases in presence of TLS}
\label{sec:impact-tls}
We proceed by grouping use cases for HTTP access faced with similar challenges.
Table~\ref{tab:use-cases} summarizes 19 use cases and evaluates the impact of HTTPS.

\begin{enumerate}[leftmargin=0pt, labelwidth=10pt, itemindent=22pt, labelsep=2pt, topsep=3pt, parsep=0pt, listparindent=10pt, itemsep=3pt]

\item \textit{Lawful interception, data retention by ISPs}.
\label{sec:lawful-intercept}
While TLS does not prevent IP traffic interception, it obviously diminishes its meaning. Protocol metadata, e.g., source and destination IP addresses, payload size, and access patterns, are still available; however, application content is not. Lawful interception of selected individuals' traffic will only yield metadata for such websites that are reached over TLS. The ISP data retention laws we looked at (in the EU and Australia) seem to only aim at listing visited websites (visible in DNS queries and the TLS SNI extension, if not encrypted~\cite{draft-ietf-tls-esni}) or simply target flows metadata, and therefore are not impacted by the switch from HTTP to HTTPS.

\item \label{sec:impact-content-filtering}\textit{Content filtering, broadcasting}.
Fine-grained inspection and techniques that require access to the web request and response content are entirely defeated by TLS. This includes content filtering (including by ISPs), IDS/IPS, AV, DLP, fraud detection, parental controls, ad blockers, transcoding and compression.
Instead, the inspection may be moved closer to the client when possible, as with browser add-ons commonly installed by AVs and for ad-blocking.
Similarly, end-to-end encrypted traffic cannot simply benefit from broadcasting; however, see~\cite{satellite-https}.

When the IPS/WAF systems are given the server's certificate private key (see details in §\ref{sec:share-priv-key}), they are able to decrypt TLS traffic up to version 1.2. In TLS 1.3, as all ciphersuites support PFS, the long-term private key is insufficient to inspect the traffic.

We make a distinction between IDS/IPS placed on the client network to monitor outbound connections to external servers, and IDS/IPS/WAF on the CDN- or server-side that analyze inbound connections.

\item \textit{DDoS}.
For L7 DDoS protections, other application-agnostic markers such as traffic provenance or uncommon traffic patterns are still relevant indicators; however, effectively distinguishing attackers from legitimate traffic benefits from the access to the cleartext traffic~\cite{rfc8404}. This is achieved as part of IPS/WAF solutions discussed above.

Volumetric DDoS protection is agnostic to TLS, and therefore does not constitute a use case for access to the HTTP traffic. However, the pre-request authorization mechanism that often comes with such protections requires the ability to answer the client's request with a challenge prior to involving the end server.

\item \textit{Load balancing and QoS}.
Assertions are often made that TLS impacts load balancing, and proposals try to address this problem~\cite{mctls,draft-camwinget-tls-use-cases-05,rfc8404}. However, in general, load balancing is not dependent on application-layer information. Typical network load balancing deals with Layer 3 (IP) or 4 (TCP). When load balancers also act as TLS terminators, the load of decrypting TLS traffic is therefore not the bottleneck (it is not yet load balanced), and hence TLS does not introduce new challenges with such load balancers.

Only application-layer load balancers (ALBs) that are not TLS terminators suffer from encrypted TLS traffic. This seems to be the case in cellular networks when MLB, traffic queuing, or dynamic QoS is used~\cite{rfc8404}.
In those three cases, TLS traffic prevents these functions from obtaining the relevant information to make decisions, possibly leading to a degraded user experience.

Cloud-based ALBs enable balancing based on the request host/path and query parameters; however, if the full requested URL is unavailable to the balancer, the application could be adapted in a way that balancing-related information appears in an outer layer, e.g., DNS or IP. The TLS SNI extension could also be used as a replacement for host-based routing.

ALBs typically communicate the source IP address to the next hop by inserting it as an HTTP header (e.g., X-Forwarded-For) when needed. Without terminating TLS sessions, such load balancers could simply encapsulate the traffic with, e.g., the PROXY protocol~\cite{proxy-protocol} that wraps TLS traffic with additional headers.

\item \textit{Tracking}.
Content-based billing is inherently hindered by TLS. Tracking of users by operators for technical reasons is equally impacted; however, such tracking does not necessarily need access to the traffic content, therefore there could be other ways of carrying the metadata information, see the above item.

\item \textit{Caching}.
When the content is encrypted, CDNs are unable to cache it. A possible delegation of unencrypted content breaks confidentiality assumptions and brings new trust issues. Other opportunistic caches done by e.g., ISPs, are hindered as well.

\item \textit{Problem resolution}.
When troubleshooting difficult~\cite{bits-ets} or application-layer~\cite{rfc8404} problems, the opaque encrypted traffic leads to inaccuracies in the diagnosis process due to the inability to locate specific transactions, user identifiers, session identifiers, URLs, and time stamps~\cite{draft-rhrd-tls-tls13-visibility-01}. This reduces the efficiency of repair services, which impacts the service availability as promised in Service Level Agreements (SLAs), and in turn incurs costs.
RFC8404~\cite{rfc8404} notes that in the absence of packet captures, application-provided diagnosis information is usually poor and unhelpful, which motivates the need to access the cleartext data.

In TLS 1.3 and TLS 1.2 with PFS ciphersuites, simply sharing a server's private key (see §\ref{sec:share-priv-key}) with troubleshooting operators does not enable them to decrypt the traffic. For short troubleshooting times at the server side, all ephemeral keys could be exported (similar to §\ref{sec:sslkeylog}).

When users suffer from an individual problem that requires targeted troubleshooting, they may be willing to participate in the effort by, e.g., running with modified configurations that are more friendly to debugging.

\item \textit{New use case: monitoring IoT devices}.
The Internet-of-Things (IoT) is a concept whereby previously unconnected objects become network-aware, are remotely controlled and interact with online services~\cite{Al-FuqahaGMAA15}. This trend is sufficiently recent for IoT devices to start encrypting their web communications from day one, unlike most other cases presented above.
The rise of consumer IoT products with multiple sensors within the user's private space sparked privacy violation concerns~\cite{RenDCMKH19}. Therefore, end users may be interested in monitoring the traffic of their IoT devices for privacy reasons. IoT traffic monitoring is a novel area that faces similar challenges as in~(\ref{sec:impact-content-filtering}). However, the main difference is that the client is no longer a browser or application of the user's choice running on a general-purpose machine, but rather a vendor-baked client on the embedded device, offering limited control.
Furthermore, IoT devices bring a unique challenge for traffic inspection as they sometimes make use of TLS client authentication using a (hardcoded) per-device certificate~\cite{akamai-iot-mutual}, a feature otherwise uncommonly used on the web.

\subhead{Summary}
TLS reduces the information extracted from network traffic to metadata only, hindering security solutions, caching, client tracking, content-based network optimization and routing decisions, legitimate modification of the traffic by third parties as well as the insertion of client challenges as part of DDoS mitigation. It also makes application and protocol errors more difficult to troubleshoot.
Considering the use cases that expect access to HTTP traffic described in §\ref{sec:http-access} and the actual impact of TLS we described here, we summarize 19 derived use cases (listed from \emph{a} to \emph{s}) for access to HTTP traffic in the presence of TLS, which remain unaddressed by alternative solutions (see Table~\ref{tab:use-cases}).
We further distilled some use cases based on the actual technical challenge, e.g., L7 DDoS prevention actually requires serving a pre-request authorization. Similarly, IDS/IPS are also split into inbound and outbound due to different technical challenges.
Notable business use cases can be mapped to several of our use cases, e.g., CDNs could span over \emph{g}, \emph{i}, \emph{l}, \emph{n} and \emph{p}.

\end{enumerate}

\newcommand{\anti}[1]{$\overline{\mbox{#1}}$}
\newcommand{\itm}[1]{\makebox[1.1em][l]{\textit{#1.}}}
\newcommand{\redno}{\textcolor{red}{\textit{No}}}

\begin{table}[h]
 \centering
 \footnotesize
 \setlength{\tabcolsep}{4pt}
 \caption{Summary of use cases (motivations) for access to plaintext traffic in the presence of TLS (from §\ref{sec:impact-tls}), the impact of HTTPS in general on them and of TLS 1.3 in particular}
 \begin{threeparttable}
  \begin{tabular}
   {%
    @{\hspace{0pt}}c@{\hspace{0pt}}|
    m{1.55in}|
    >{\centering}m{.75in}|
    >{\centering}m{.85in}|
    >{\centering}m{.55in}|
    m{.7in}|
    m{.95in}|
   }
   \multicolumn{1}{c}{~} & \bfseries{Use cases} &
   \multicolumn{1}{m{.8in}|}{\bfseries{Can be conducted with HTTPS traffic}} &
   \multicolumn{1}{m{.85in}|}{\bfseries{Effectiveness with HTTPS \newline (vs.\ HTTP traffic)}} &
   \multicolumn{1}{m{.55in}|}{\bfseries{Impact of TLS 1.3 \newline (vs.~TLS~1.2)}} &
   \bfseries{Related \newline operations} &
   \bfseries{Immediate \newline alternatives}
   \\ \noalign{\hrule height 1pt}
   
   \multirow{6.3}{*}{\rotatebox{90}{Legal}} &\itm{a} Lawful interception &
    Yes, \textcolor{red}{\textit{limited}}    &
    \textcolor{red}{\textit{Metadata only}}   &
    \textcolor{red}{\textit{Low}}$^\dagger$    &
    --    &
    --  \\
   
   \cline{4-4}
   ~& \itm{b} ISP data retention &
    Yes, \textcolor{red}{\textit{limited}}    &
    \textcolor{red}{\textit{Limited browsing history}}    &
    None    &
    --    &
    --   \\
   
   \cline{4-4}
   \cline{7-7}
   ~&\itm{c} Legal content filtering &
    Yes, \textcolor{red}{\textit{limited}}    &
    \textcolor{red}{\textit{Less granularity}}    &
    None    &
    --    &
    DNS filtering, \newline{}IP blocking  \\
   
   \cline{6-7}
   ~&\itm{d} Fraud detection &
    \redno    &
    \textcolor{red}{\textit{None}}    &
    None   &
    Employee\newline{}monitoring    &    
    Endpoint \newline monitoring  \\
    
   \hline
   
   \multirow{16.3}{*}{\rotatebox{90}{Security/privacy}} & \itm{e} Malware download prevention &
    Yes, \textcolor{red}{\textit{limited}}    &
    \textcolor{red}{\textit{Domain blocking only}}   &
    None  &
    --   &
    Browser addon  \\
   
   \cline{4-4}
   \cline{6-7}
   ~& \itm{f} Outbound IDS/IPS &
    \redno   &
    \textcolor{red}{\textit{None}}   &
    None  &
    DLP (see \emph{j})  &
    Endpoint \newline monitoring \\
   
   \cline{5-7}
   ~& \itm{g} Inbound IDS/IPS/WAF &
    \redno    &
    \textcolor{red}{\textit{None}}    &
    \textcolor{red}{\textit{No key sharing}}   &
    L7 DDoS \newline prevention,\newline  DLP (see \emph{j}) &
    Earlier TLS \newline termination \\
   
   \cline{5-7}
   ~& \itm{h} Data retention (compliance) &
    Yes, \textcolor{red}{\textit{limited}}    &
    \textcolor{red}{\textit{Metadata only}}    &
    \textcolor{red}{\textit{Low}}$^\dagger$    &
    Attack \newline investigation, forensics   &
    Endpoint monitoring, earlier TLS \newline termination   \\
    
   \cline{6-7}
   ~& \itm{i} Pre-request authorization &
    \redno    &
    \textcolor{red}{\textit{None}}    &
    None    &
    L7 DDoS \newline prevention,\newline{}Turing test &
    --   \\
    
   \cline{4-7}
   ~& \itm{j} Data loss prevention (DLP) &
    Yes, \textcolor{red}{\textit{limited}}    &
    \textcolor{red}{\textit{Metadata only}}    &
    None    &
    IPS (see \emph{f} and \emph{g})   &
    Endpoint/server \newline monitoring  \\
   
   \cline{4-7}
   ~& \itm{k} IoT device monitoring &
    Yes, \textcolor{red}{\textit{limited}}    &
    \textcolor{red}{\textit{Metadata only}}   &
    None  &
    --    &
    --    \\
    
   \hline
   
   \multirow{5.3}{*}{\rotatebox{90}{Performance}} & \itm{l} L7 load-balancing &
    \redno    &
    \textcolor{red}{\textit{None}}    &
    None    &
    --    &
    --    \\
   
   ~& \itm{m} Broadcasting &
    \redno    &
    \textcolor{red}{\textit{None}}    &
    None    &
    --    &
    --    \\
   
   ~& \itm{n} Server-mandated caching &
    \redno    &
    \textcolor{red}{\textit{None}}    &
    None    &
    CDN    &
    --    \\
  
   ~& \itm{o} Opportunistic caching &
    \redno     &
    \textcolor{red}{\textit{None}}     &
    None     &
    ISP caching    &
    --     \\
  
   ~& \itm{p} Transcoding/compression &
    \redno     &
    \textcolor{red}{\textit{None}}     &
    None     &
    --     &
    --   \\
    
   \hline
   
   \multirow{4.3}{*}{\rotatebox{90}{Other}} & \itm{q} Problem troubleshooting \newline \itm{q1} ...with user help \newline \itm{q2} ...with no user help &
    Yes, \textcolor{red}{\textit{limited}}    &
    \textcolor{red}{\textit{Metadata only}}    &
    \textcolor{red}{\textit{No key sharing}}    &
    --    &
    Endpoint monitoring, application logging   \\
   
   \cline{2-7}
   ~& \itm{r} Parental control &
    Yes, \textcolor{red}{\textit{limited}}    &
    \textcolor{red}{\textit{Less granularity}}    &
    None    &
    --    &
    Browser addon,\newline{}special browser  \\
    
   \cline{7-7}
   ~& \itm{s} ISP billing/tracking &
    \redno   &
    \textcolor{red}{\textit{None}}    &
    None    &
    --    &
    --     \\
    
   \hline
   
  \end{tabular}
  \begin{tablenotes}[flushleft]\footnotesize 
   \item[]\textcolor{red}{\textit{Red color}} denotes a negative impact for the third parties that want plaintext access.
   Note that the second column does not consider certificate private key sharing or TLS session splitting as a way to conduct operations, although they are common technical practices in some cases. However, we consider the impact of TLS 1.3 on such practices as there is otherwise no other impact (fourth column).
   In the third column, metadata refers to e.g., domain information, traffic patterns, payload size.
   $^\dagger$Low impact: the third party does not learn the server certificate, and connections are forward-secure (which is not always the case with TLS 1.2).
  \end{tablenotes}
 \end{threeparttable}
 \label{tab:use-cases}
\vspace{-.05in}
\end{table}

\subsection{Industry regulations}
\label{sec:industry-reg}
Industry regulations may seem to require access to plaintext network traffic or prohibit its inspection, depending on the objective.
We discuss the limited arguments stating that some regulations require access to plaintext traffic and would be negatively impacted by TLS.

An Internet Draft by Cisco~\cite{draft-camwinget-tls-use-cases-05} argues that industry regulations such as NERC Critical Infrastructure Protection (CIP) and the Payment Card Industry Data Security Standard (PCI-DSS) are impacted by TLS 1.3.
However, we are unable to find documented evidence to support this claim.
For instance, the authors point that a PCI-DSS requirement ``\emph{describes the need to be able to detect protocol and protocol usage correctness.}'' We could not locate this specific requirement in the referenced version of the standard.
The closest match would be Requirement 4.1.c which states: ``\emph{Select and observe a sample of inbound and outbound transmissions as they occur (for example, by observing system processes or network traffic) to verify that all cardholder data is encrypted with strong cryptography during transit.}''
The stated verification could be achieved by asserting that: 1) no cardholder data is sent unencrypted, and that 2) when encrypted connections are detected, the negotiated ciphersuite is compliant, which is not hindered by TLS 1.3.
The draft also points to Requirement 10 about ``\emph{the need to provide network-based audit to ensure that the protocols and configurations are properly used}''~\cite{draft-camwinget-tls-use-cases-05}. We are unable to find support for this claim from reading the corresponding PCI-DSS standard requirement, and were unable to contact the draft's authors for clarification.

Similarly, we could not identify requirements in NERC CIP impacted by TLS, especially version 1.3.
Overall, to our knowledge, there is no major industry standards that are affected by HTTPS compared to plain HTTP and specifically by TLS 1.3. If such conflicts actually exist, we urge industry regulators to document the problem to enable the community to find solutions.

\section{Stakeholder incentives}
\label{sec:stakeholders}
We now perform an incentive analysis from the viewpoint of the main stakeholders.
Since the access by a third party to the plaintext content of TLS communications is normally considered an attack on the protocol, at least one TLS endpoint needs to collude with the third party to allow it to happen. That is, without the collaboration of a TLS endpoint (therefore of the end user, its client or the server), these use cases are prohibited. We explore below the various incentives of multiple stakeholders for this collaboration, and list these incentives with respect to the 19 use cases in Table~\ref{tab:stakes}.
We consider the stakeholders' incentives while evaluating proposals in §\ref{sec:evaluation}. 

\begin{table}[t!]
 \centering
 \small
 \setlength{\tabcolsep}{4pt}
 \caption{Incentives and requirements for the main stakeholders in each use case for access to plaintext traffic in the presence of TLS summarized from §\ref{sec:impact-tls}.}
 \vspace{-.2in}
 \begin{threeparttable}
  \begin{tabular}
   {%
    @{\hspace{0pt}}c@{\hspace{0pt}}|
    m{2in}|
    >{\centering}m{.3in}|
    >{\centering}m{.3in}|
    >{\centering}m{.3in}|
    >{\centering}m{.3in}|
    >{\centering}m{.3in}|
    c|
   }
   \multicolumn{2}{p{50mm}}{\diagbox[width=56mm]{\bfseries{~Use cases}}{\bfseries{Stakeholders}}} &
   \mcrot{1}{l}{45}{{End user$\dagger$ }} &
   \mcrot{1}{l}{45}{{Client vendor$\dagger$}} &
   \mcrot{1}{l}{45}{{End-user organization}} &
   \mcrot{1}{l}{45}{{ISP}} &
   \mcrot{1}{l}{45}{{CDN}} &
   \mcrot{1}{l}{45}{{{Content provider}}}
   \\ \noalign{\hrule height 1pt}
   
   \multirow{4.3}{*}{\rotatebox{90}{Legal}} &\itm{a} Lawful interception &
    $\star$    &
    ~    &
    ~    &
    {L}    &
    ~    &
    ~    \\
    
   ~& \itm{b} ISP data retention &
    $\star$    &
    ~    &
    ~    &
    {L}    &
    ~    &
    ~    \\
    
   ~&\itm{c} Legal content filtering &
    $\star$    &
    ~    &
    ~    &
    {L}    &
    ~    &
    ~    \\
    
   ~&\itm{d} Fraud detection &
    e$\star$    &
    ~    &
    {LE}   &
    ~    &    
    ~    &
    ~    \\
    
   \hline
   
   \multirow{7.3}{*}{\rotatebox{90}{Security/privacy}} & \itm{e} Malware download prevention &
    {SL}   &
    S    &
    {SLE}  &
    e   &
    e   &
    e$\star$   \\
    
   ~& \itm{f} Outbound IDS/IPS &
    sl$\star$   &
    s$\star$   &
    {SLE}  &
    ~    &
    ~    &
    $\star$    \\
    
   ~& \itm{g} Inbound IDS/IPS/WAF &
    ~    &
    ~    &
    ~    &
    ~    &
    {SE}   &
    {SLE$\star$} \\

   ~& \itm{h} Data retention (compliance) &
    ~    &
    ~    &
    {E}    &
    ~    &
    ~    &
    {E}    \\
    
   ~& \itm{i} Pre-request authorization &
    $\star$    &
    ~    &
    ~    &
    ~    &
    {SE}   &
    {SE}   \\
    
   ~& \itm{j} Data loss prevention &
    $\star$    &
    $\star$    &
    {SLE}  &
    ~    &
    ~    &
    {SLE}  \\
    
   ~& \itm{k} IoT device monitoring &
    {S}    &
    se$\star$   &
    ~    &
    ~    &
    ~    &
    se \\
    
   \hline
   
   \multirow{5.3}{*}{\rotatebox{90}{Performance}} & \itm{l} L7 load-balancing &
    f   &
    ~    &
    ~    &
    {FE}   &
    {FE}   &
    {FE}   \\
    
   ~& \itm{m} Broadcasting &
    f   &
    ~    &
    ~    &
    {FE}   &
    e   &
    e   \\
    
   ~& \itm{n} Server-mandated caching &
    f   &
    f   &
    ~    &
    ~    &
    FE   &
    {FE}   \\
    
   ~& \itm{o} Opportunistic caching &
    ~    &
    ~    &
    {FE}   &
    {FE}   &
    ~    &
    ~    \\
    
   ~& \itm{p} Transcoding/compression &
    f$\star$   &
    ~    &
    ~    &
    {FE}   &
    {FE}   &
    f   \\
    
   \hline
   
   \multirow{3.3}{*}{\rotatebox{90}{Other}} & \itm{q} Problem troubleshooting &
    f   &
    ~    &
    ~    &
    ~    &
    ~    &
    {FE}   \\
    
   ~& \itm{r} Parental control &
    {S$\star$}    &
    S    &
    ~    &
    E    &
    ~    &
    ~    \\
    
   ~& \itm{s} ISP billing/tracking &
    $\star$    &
    ~    &
    ~    &
    {FE}   &
    ~    &
    ~    \\
    
   \hline
   
  \end{tabular}
  \begin{tablenotes}[flushleft]\footnotesize
   \item[] Categories of stakeholder incentives and requirements for access to plaintext traffic: legal (L), security and/or privacy (S), performance (F), and economic/business (E).
   We list the primary incentives/requirements under each stakeholder using capital letters (L,S,F,E), and denote by the corresponding lowercase letters (l,s,f,e) when a stakeholder could reasonably offer some help (if not too costly) to support the preceding reasons.
   A star indicates stakeholders that are opposed to plaintext access for that use case ($\star$). 
   $\dagger$ denotes an assumption of a non-malicious stakeholder.
  \end{tablenotes}
 \end{threeparttable}
 \label{tab:stakes}
\end{table}

\subhead{Examples of incentives}
These use cases are associated with different stakeholders with various degrees of interest. For instance, the end user is unrelated to data retention compliance requirements faced by the server operator, and could even be adverse to their ISP inserting tracking headers (see~\cite{verizon-supercookie}). Likewise, the server entities are unlikely interested in how companies prevent data leakages from their network. However, several parties may have an interest to detect malware downloads: the end user or enterprise network administrators in the first place for security reasons; the client vendor is usually interested in user safety (see e.g., Google Safe Browsing~\cite{safe-browsing}); the ISP could also gain business advantages by proposing a ``more secure'' Internet access to its users; finally, the server operators could choose to be compliant with solutions that assert their traffic is benign to enhance their public image.

\subhead{Stakeholders}
For the sake of our analysis, we consider five main stakeholders: the end user (EU), client vendor/author (CV, e.g., browser vendor), end-user organization (EUO), ISP, CDN, and content provider (CP).
However, note that the EUO may also outsource security operations on their traffic to cloud-based companies (see~\cite{embark}). In this case, the organization may want to enable the cloud-based solution to inspect TLS traffic, yet may be reluctant to fully trust the provider not to take advantage of the unencrypted data. This situation sparks a separate problem in which the EUO is split into two entities.

\subhead{Categories of incentives}
\label{sec:cat-incentives}
We identify four categories of stakeholder incentives and requirements for access to plaintext traffic: legal, security and/or privacy, performance, and economic/business. 
For each use case for access to plaintext traffic in the presence of TLS and for each stakeholder, we list in Table~\ref{tab:stakes} the corresponding incentives/requirements.
A stakeholder may not always represent entities that share the same incentives. For instance, the author of a piece of malware acting as a TLS client may not willingly support functions that would defeat its intended purposes, the way browser vendors would. Similarly, the corresponding attacker-controlled server will not support features that would, e.g., facilitate intrusion detection in the victim's network.

\subhead{Influence of stakeholders}
Note that the EUO may be able to impose its incentives onto the user browsers and EU as seen in organizations that control user devices and enforce strict application whitelisting (e.g., via Mobile Device Management solutions).
End users can also influence browser vendors to support a feature they want.

\subhead{No mutual agreement}
Both the EU/CV/EUO and the CP never share a strong incentive to support a use case among the 19 considered. Note that for data retention for compliance (\emph{h}) and data loss prevention (\emph{j}), the incentives of both sides are disjoint. This observation is particularly relevant when designing solutions. Indeed, a proposal that requires the collaboration of both ends is unlikely to be supported.

\section{Session splitting and key sharing}
\label{sec:enterprise-twists}
Faced with immediate road blocks imposed by TLS, network operators' first recourse are simple solutions that bypass TLS by splitting TLS sessions, or sharing TLS keys/secrets once established. These solutions only require the middlebox and a single endpoint to be compatible, while the other endpoint remains largely unaffected. In this section and the following ones, we list prominent issues for each technique and proposal discussed.

\subsection{TLS session splitting, a.k.a.\ man-in-the-middle}
\label{sec:split-sessions}
The simplest solution to access the unencrypted traffic of an outbound connection is for a middlebox to interpose the TLS connection and \emph{split} it into two sessions~\cite{transitive-trust-dell} (not to be confused with the encryption/authentication splitting described in~\cite{Lesniewski-LaasK05}). This behavior is often (abusively) referred to as a man-in-the-middle attack.
The client is directed to the middlebox acting as a server, which in turn acts as a client and exchanges with the intended server. Hence, the middlebox has two sides: client-side and server-side, respectively.
The middlebox also acts as a CA with its own root certificate since it cannot simply impersonate the server using publicly trusted certificates. This method requires proper configuration of the client to accept middlebox-issued certificates. 
The middlebox is trusted to perform the client validations (e.g., certificate validation, minimum key length, strong signature hashing algorithm), and to maintain an acceptable level of security on the server-side connection (e.g., same TLS version and ciphersuites).
This method is often adopted by antivirus, parental control, malware, student/employee monitoring systems, anti-ads software, and enterprise network appliances~\cite{transitive-trust-dell,xcc-tls-proxy,https-intercept-impact,WakedMY18,carnavalet-wajam}.

When the middlebox leverages a publicly trusted intermediate CA certificate, clients do not require the extra provisioning of the middlebox' root certificate. However, clients are mostly unaware that the interception is taking place, and OS/browser vendors sanctioned the issuing CAs when this deceptive practice was discovered (see~\cite{trustwave-revokation,turktrust-revokation,anssi-revokation,cnnic-revokation}).
Thus, splitting TLS sessions is only relevant when considering home/enterprise networks.

\subhead{Issues}
The client does not see the server's certificate; thus this simple technique precludes any certificate-based verification and enforcement by the client, e.g., certificate validation, revocation checking, key pinning~\cite{rfc7469}, DANE~\cite{rfc7671}. Instead, the middlebox is trusted to perform such operations consistently with the client's expectations.
Also, the server's use of Extended Validation (EV) certificates is downgraded to regular Domain Validation (DV) certificates; however, note that EV certificates for the web are no longer differentiated by major browsers~\cite{chrome-ev,firefox-ev}.

In practice, TLS session splitting middleboxes have been shown to overlook critical steps as well as introduce additional attack surface~\cite{xcc-tls-proxy,kaspersky-32bit,https-intercept-impact,WakedMY18}.
This approach is also incompatible with TLS client authentication unless the middlebox authenticates on behalf of the client to the server.

Finally, it is not always easy or even possible to provision clients to trust a custom root CA. Specifically, user devices that are not controlled by the organization that performs the session splitting needs to be provisioned by the user. Certain clients such as antivirus applications or IoT devices rely on their own trust store with little support for customization~\cite{xcc-tls-proxy}.

\subheadit{ProxyInfoExtension}
An Internet draft~\cite{draft-mcgrew-tls-proxy-server-01} proposes that the middlebox carries the server's certificate back to the client in a new informational TLS extension, named \textsf{ProxyInfoExtension}, as part of the \textsf{ServerHello} to the client. The client is then responsible for performing validity checks and decide to continue or not the connection.
The middlebox is trusted to place the actual server certificate in this extension, as there is no cryptographic binding between this certificate and the rest of the server-facing connection. Although incomplete to address all criticisms, this technique aims to take the middlebox out of certificate-related decisions, reducing potential security issues and bringing more transparency to the client.

\subsection{Certificate private key sharing}
\label{sec:share-priv-key}
When in possession of the server certificate's corresponding private key, two options arise.

\subsubsection{Passive decryption using RSA certificates}
When TLS endpoints negotiate an RSA key exchange ciphersuite, and thus the server presents an RSA certificate, the ePMS can be decrypted by any entity that possesses the certificate private key.
Therefore, sharing the server's certificate private key is an effective solution to passively monitor recorded TLS traffic by, e.g., IDS~\cite{ssl-ids,draft-green-tls-static-dh-in-tls13-01,draft-rhrd-tls-tls13-visibility-01}.

\subhead{Issues} This technique for passive monitoring is incompatible with DH key exchange ciphersuites, including TLS 1.2 PFS ciphersuites and TLS 1.3 altogether.

\subsubsection{Active impersonation}
\label{sec:active-impersonation}
The private key of an RSA or EC certificate can be shared for active inspection or impersonation of the server (e.g., content caching) as seen with several commercial CDNs~\cite{https-meets-cdn}.
This technique falls back to TLS session splitting (§\ref{sec:split-sessions}), but using the server certificate.

\subhead{Issues} This technique and the previous one question the notion of ``private'' key that is no longer known to one entity only. More importantly, impersonating the final server (e.g., on a TLS-terminating CDN node, i.e., edge node) also challenges the notion of end-to-end encryption when the server-end may not be the actual end, and without the knowledge of the end user.

\subsection{Sharing DH key exchange private shares}
\label{sec:dh-key-share}
We first describe the general technique of sharing DH secrets, then discuss protocols that rely on it.
\subsubsection{Static DH key sharing}
\label{sec:dh-key-share-vanilla}
With (EC)DHE key exchanges, the session secret is only known to the two entities that know a private part of the key. However, unlike with certificate sharing, new (EC)DHE keys are generated for each new session on both ends (i.e., they are ephemeral keys), which precludes pre-sharing a single key.
Therefore, to facilitate traffic monitoring at the server-side, a simple solution is to make the server's DH key share static, then share the private part with a middlebox just as one would share an RSA certificate private key (§\ref{sec:share-priv-key}). The TLS session secrets are reconstructed by the middlebox in the same way as the server does. Since RSA key exchange and static DH ciphersuites were dropped in TLS 1.3, this method comes as an attractive solution to re-enable passive monitoring in a way that is similar to the well-known certificate private key sharing.

\subhead{Issues} The resulting semi-static DH key exchange no longer provides forward secrecy.
Similar to §\ref{sec:share-priv-key}, a private key share is no longer private to the end entity. However, server impersonation requires the active participation of the server to sign the \textsf{CertificateVerify} message. Therefore in this case, other alternatives are more suitable (see Keyless SSL~§\ref{sec:keyless-ssl}).
Moreover, if the client is unaware that the server uses a static key and also decides to choose a static key for its own needs (e.g., client-side passive monitoring), the resulting shared secret remains the same across sessions.
The uniqueness of the derived TLS secrets then only depends on the client and server random values. If those values suffer from implementation flaws, plaintext communications may even be recovered~\cite{redhat-dtls-random}.
This situation could be avoided by making explicit that one party uses a static share.

\subsubsection{Enterprise Transport Security}

\label{sec:ets}
ETS~\cite{ets} (formerly known as eTLS) is a variant of TLS 1.3, promoted by the Bank Policy Institute and standardized by the ETSI. It combines TLS splitting, server impersonation, and static DH keys on middleboxes/firewalls or enterprise servers to passively decrypt traffic within an enterprise. ETS addresses three scenarios:

\begin{enumerate}[leftmargin=0pt, labelwidth=28pt, itemindent=48pt, labelsep=2pt, topsep=1pt, parsep=0pt, listparindent=10pt, itemsep=3pt, label=Scenario~\arabic*.]
\item Both clients and servers reside in the enterprise. The servers share static DH keys with passive middleboxes. Client-server connections do not benefit from PFS.
\item The clients are located outside the enterprise network. The clients' TLS connections are first terminated at a firewall that impersonates the server as described in §\ref{sec:active-impersonation}, then are re-encrypted within the enterprise up to the final server that employs a static DH key.
This scenario is similar to the one described in an Internet Draft~\cite{draft-green-tls-static-dh-in-tls13-01} in the context of monitoring re-encrypted traffic within datacenters, after the client-originating TLS traffic is terminated at a load balancer.
Note that PFS can be preserved on the client-firewall or client-datacenter segment using regular TLS stacks.

\item Enterprise clients try to reach servers outside the enterprise network. A firewall at the edge of the network impersonates the end servers by splitting TLS sessions as in §\ref{sec:split-sessions}, albeit with static DH keys for other middleboxes to decrypt the traffic.
\end{enumerate}

\subhead{Criticisms}
ETS received vivid criticisms due to its weakening of TLS 1.3. It has been assigned CVE-2019-9191 for lacking forward secrecy.
An Internet draft by the American Civil Liberties Union (ACLU)~\cite{draft-dkg-tls-reject-static-dh-01} also proposes that clients reject connections to servers/middleboxes implementing ETS, while the Electronic Frontier Foundation (EFF) calls for ETS not to be used~\cite{ets-eff}. A request from ETSI to NIST aiming at delaying the publication of TLS configuration guidelines to include the near-finalized ETS standard was dismissed~\cite{nist-etsi-ets}.

We note that in the three scenarios described in the ETS standard, TLS clients outside the enterprise never reach an ETS server directly. Rather, those client connections end at the edge firewall/load balancer that acts as the final TLS server with support for forward secrecy. In other words, the part of the connections that goes on the public Internet is protected by vanilla TLS. Once inside the enterprise or datacenter, the traffic is relayed to the final server over ETS, i.e., TLS with semi-static DH keys. Thus, the accusations against ETS seem to be partially ill addressed. Note that while end users are unaware that their traffic is partially carried over ETS, they would also not know if it was done in plaintext (see Google's scenario described in NSA's MUSCULAR program~\cite{google-nsa-ssl}). This issue relates to the shift of the perceived ``server-end'' from an application server to an edge server.

\subhead{Issues}
ETS describes a key manager server that can either push or serve keys to enterprise servers and middleboxes. This aspect of the protocol aims to simplify the provisioning of keys.
However, it significantly complicates deployments by adding an online key server that should properly authenticate the requests for keys. In turn, such new interfaces increase the attack surface.
Also, ETS binds the TLS certificate to a hash of the static DH public key through a certificate extension. Thus, the validity period of certificates for ETS implicitly matches the use period of the DH key. In turn, those certificates should be rotated as well.
In all three scenarios, ETS is or can be coupled with enterprise-trusted CA certificates, so the certificate rotation does not necessarily involve an external CA, which could otherwise pose as a bottleneck.

\subsubsection{tls\_visibility}
\label{sec:tls-visibility}
An Internet draft~\cite{draft-rhrd-tls-tls13-visibility-01} suggests a mechanism in TLS 1.3 for connections to be inspected by a pre-approved third party. Clients opt-in by advertising their willingness in the \textsf{ClientHello}. This proposal is suitable for inspecting TLS connections by enterprise middleboxes to a server within the enterprise or datacenter.
The server is given a long-term ECDH public key \textsf{DH$_1^+$} while authorized middleboxes are given the private counterpart \textsf{DH$_1^-$}. Upon an incoming TLS 1.3 connection with the empty \textsf{tls\_visibility} extension present, the server generates a short-term ECDH key pair \textsf{DH$_2^+$}/\textsf{DH$_2^-$}, independently of the TLS key exchange (EC)DHE key pair. Then, the server calculates the shared secret $Z$ as the result of the key exchange involving \textsf{DH$_1^+$} and \textsf{DH$_2^-$} (requiring both keys to be selected from the same elliptic curve).
The server responds with a \textsf{ServerHello} that includes the \textsf{tls\_visibility} extension composed of \textsf{DH$_2^+$} along with the \textsf{early\_secret} and \textsf{handshake\_secret} encrypted with a key $Ke$ derived from $Z$. Middleboxes are able to reconstruct $Z$ using the provisioned \textsf{DH$_1^-$} and the given \textsf{DH$_2^+$} from the \textsf{ServerHello}. Thereafter, the middleboxes can decrypt the TLS secrets, and derive the keys to decrypt the rest of the traffic.

\subhead{Issues}
Malicious users can simply short-circuit an IPS that relies on the client's opt-in to receive the TLS secrets.

\subsection{Client session key sharing}
When splitting the TLS session is unacceptable at the client-side, some techniques and proposals similarly target TLS secret sharing at the client-side.

\subsubsection{SSLKEYLOGFILE}
\label{sec:sslkeylog}
Some browsers (e.g., Chrome, Firefox) can be asked to write TLS secrets to a log file~\cite{sslkeylog-nss} (e.g., PMS, MS, or handshake/application traffic secrets for TLS 1.3), which allow for passive decryption of the corresponding TLS traffic. This mechanism is generally enabled by setting the SSLKEYLOGFILE environment variable to a writable path. This approach has been adopted by some antivirus applications~\cite{avast-sslkeylog} by pointing the environment variable of browser processes to, e.g., a Windows named pipe.
A similar approach for unsupported applications consists in instrumenting the application's TLS library to extract the TLS secrets, see~\cite{BatesPNHTBA14,peetch}.

\subhead{Key exfiltration}
By leveraging SSLKEYLOGFILE or by modifying the TLS implementation, keys can then be exfiltrated to a middlebox to provide filtering capabilities without splitting the TLS session, see LOCKS~\cite{locks}.
Commercial solutions for cloud environments also propose to scan a virtual machine memory upon finishing a TLS handshake to recover and aggregate session keys without application modification~\cite{nubeva-tls-decrypt}.

\subhead{Issues}
Support for this feature is limited to few applications and restricted (e.g., user warning, or special builds~\cite{sslkeylog-nss}) due to the obvious potential for misuse, i.e., this could enable easier and less privileged traffic-intercepting malware. Similar to DH key sharing on the server-side (§\ref{sec:dh-key-share-vanilla}), the key exfiltration requires a separate infrastructure. This is especially problematic on the client-side, arguably more hostile than a controlled server environment.

\subsubsection{TLS-RaR}
\label{sec:tls-rar}
For IoT traffic monitoring, TLS-RaR~\cite{tls-rar} proposes to release TLS session keys for communications once they are finished. This schedule preserves the integrity of IoT traffic while making it auditable. This compromise may stimulate IoT vendors to open their opaque traffic.

\subhead{Issues}
Until the device releases past session keys, the end user has no guarantee that all the current traffic will be auditable as there is no commitment or escrow mechanism.

\section{Delegation of content and keys}
\label{sec:cdn}
An alternative set of approaches tailored for caching middleboxes (e.g., CDNs) delegates parts or all of the content delivery operations to third party-owned middleboxes in a way that they do not learn of the server's long-term key.
This can be done with a varying degree of control by the server or trust in the middlebox. By preventing the middlebox from distributing unapproved content, the server can retain full control. The server can also act as an online oracle for private key operations and give more freedom to the middlebox. Finally, the server can ``vet'' for middleboxes and remove itself from the communication for some duration.
Note that in the CDN context, the end-server is referred to as the origin server, while the CDN-owned node is called the edge node.

\subsection{Server-controlled caching}
\label{sec:enc-mac-splitting}
We discuss below two main approaches for caching content with the server's collaboration and control.

\subsubsection{Proxy encrypts, server authenticates}
\label{sec:untrusted-caches}
Lesniewski-Laas and Kaashoek~\cite{Lesniewski-LaasK05} propose to ``split'' the encryption and authentication layers in TLS to allow untrusted collaborative proxies to distribute authenticated cached contents without the knowledge of the server's private key in a solution called Barnraising. The proxies are willingly participating, and do not need to be trusted to deliver the correct content.
The protocol relies on the disclosure of the \texttt{server\_write\_key} by the origin server to the caching proxy, 
which enables the proxy to encrypt the cached content on behalf of the server. The missing authentication tags to form valid TLS records are provided by the server, which also encrypts the requested content as if it was fulfilling the request by itself; however, it only sends the corresponding tags to the proxy.
From the client's perspective, there is no difference from a direct TLS session with the origin server.

The encryption layer is actually superfluous for the target goal. In fact, the authors point that an authentication-only TLS ciphersuite could be used; however, this would require the client to advertise such ciphersuite, which in practice is not the case for modern browsers, and is deprecated in TLS 1.3.

A patent by Akamai~\cite{gero2015splicing} proposes a similar idea aimed at selectively enabling a CDN to serve a request from its cache, without the knowledge of the server's private key.
The main difference with Barnraising resides in the ability of
the origin server to decide when to share the \texttt{client\_write\_key} to enable the edge server to decrypt client requests.
If the edge server deems that it can serve the requested content from cache, it informs the origin server, which then shares the \texttt{server\_write\_key} and authentication tags as described above.

The patent also expresses the natural revocation of the write capability of the edge server since the origin can simply stop providing further MACs.
The read capability allegedly can be revoked by requesting the client to renegotiate, effectively establishing new keys that the middlebox no longer knows. However, this mechanism seems insufficient to prevent a malicious middlebox from giving up read capability from the client, and read and write capabilities from/to the server.\footnote{When the edge server knows the \texttt{client\_write\_key}, it can renegotiate with the origin server without the client being involved, as it can decrypt \texttt{verify\_data} from the client's \textsf{Finished} message, even with the secure renegotiation extension~\cite{rfc5746}.
The edge server can then still decrypt client requests (with the former \texttt{client\_write\_key}), and forge requests to the origin server and decrypt the response.}

\subhead{Issues}
This scheme only works for non-AEAD ciphersuites since the encryption and MAC keys are separate. With AEAD, authentication tags are generated during encryption, allowing the edge server to impersonate the origin server.
Thus, PFS ciphersuites in TLS 1.2 are not supported, and the scheme is incompatible with TLS 1.3.
Also, caching middleboxes only save network bandwidth from the origin, not computing resources.

\subsubsection{Public/private content splitting}
\label{sec:qos23}
QoS3~\cite{al2019qos3} proposes to establish two connections to the server. One is end-to-end encrypted and delivers the main \emph{private} document in which \emph{public} content is tagged by the server (e.g., as an HTML tag attribute). Then, the public content can be fetched explicitly by browsers using a second proxied HTTPS connection that is intercepted by a caching middlebox. The integrity of the content is verified using pre-shared signatures from the main connection, or by leveraging the Subresource Integrity (SRI) mechanism~\cite{sri}.
This proposal improves on QoS2~\cite{qos2} whereby the delegated connection is plain HTTP, a design justified by the fact that public resources may not need confidentiality guarantees. However, plaintext connections to the server are incompatible with HSTS policies.

\subhead{Issues}
The browser should withhold cookies from being sent over the proxied connection to avoid exposure of private content. Also, the solution only addresses caching for content served on the same domain.

\subsection{Server-authenticated caching (per-session content delegation)}
\label{sec:server-authd-caching}
The content delegation schemes below let the CDN decide on the content to serve; however, the origin server is still involved in the authentication phase (signing/decrypting) for each TLS session.

\subsubsection{Keyless SSL}
\label{sec:keyless-ssl}
The Cloudflare CDN offers Keyless SSL~\cite{keyless-ssl,StebilaS15}, a solution to serve content over HTTPS without the knowledge of a customer's private key by requiring that the origin server also act as a key server that can sign and decrypt a key share or ePMS (for DHE and RSA key exchanges, resp.) upon request by the edge server.
The latter is then able to compute the shared secret and derive symmetric encryption keys with the client.

\subhead{Protocol flaws}
The key server remains agnostic to TLS and simply acts as a signing/decryption oracle, communicating over a mutually authenticated channel with the edge server. In this setup, Keyless SSL was found to be vulnerable to two attacks when an attacker compromises an edge server~\cite{keyless-ssl-analysis}. First, for RSA handshakes, an attacker can query the origin key server from a compromised edge to decrypt previously captured ePMS, which leads to the full decryption of the corresponding traffic. Second, for DHE handshakes, the attacker can request a signature on a Server Config (SCFG) message as used in the QUIC protocol. The malicious SCFG message contains the attacker's DH parameters and a large expiration time. The attacker can then act as a QUIC server and impersonate the origin server to clients until the origin server certificate is revoked. Although these attacks have been published in 2017, there is evidence that Cloudflare was aware of at least the first one in 2012~\cite{smith-sullivan-tweet}.

Bhargavan et al.~\cite{keyless-ssl-analysis} propose to fix these issues by making the key server aware of TLS in a solution called 3(S)ACCE-K-SSL. In particular, it is the key server that generates and signs DHE parameters, and encrypts the \textsf{Finished} message given the transcript of the handshake between the client and the edge server. The edge does not know the master secret and is unable to resume the session by itself. The authors also discuss the need for an extra PKI to distinguish middleboxes and authenticate each piece of content to be served. This solution requires three round trips with the origin server instead of one, thus comes at a significant performance cost.

\subhead{Other issues}
One drawback of Keyless SSL is that the key server must be highly available, partially defeating the purpose of using a CDN, and adding a round trip between the edge and the origin servers per TLS handshake.

\subsubsection{LURK}
\label{sec:lurk}
Limited Usage of Remote Key (LURK~\cite{draft-mglt-lurk-lurk-00,BoureanuMPAMFM20}) is a general protocol for context-dependent remote interactions with cryptographic material. The goal is to offload sensitive key material to a LURK server, and let LURK clients interact with it for cryptographic operations, e.g., during a TLS handshake. Contexts are described in LURK extensions (see for TLS~\cite{draft-mglt-lurk-tls12-05,draft-mglt-lurk-tls13-05}).
LURK is positioned as an alternative to Keyless SSL, thus it is also tailored for the CDN use case; however, the fundamental idea remains the same: the key server either decrypts an ePMS or signs DHE parameters. The standard defines a protocol between LURK clients (e.g., CDN edges) and servers (i.e., key server), based on a mutually authenticated secure channel. The authors formally verified the security guarantees of LURK for TLS 1.2 with ProVerif.

The two major attacks on Keyless SSL identified by Bhargavan et al.~\cite{keyless-ssl-analysis} are addressed differently than prescribed by the authors. LURK lets the edge server pick a pre-\texttt{server\_random} value $S$. The key server then derives \texttt{server\_random}=$\varphi_k(S)$ before decrypting the ePMS or signing the DHE parameters. $\varphi$ is called a freshness function, it is a PRF keyed with a key exported from the secure channel (see~\cite{rfc5705}). The derivation prevents an attacker from obtaining a signature on a replayed \texttt{server\_random}, as the attacker is unaware of $S$. Even in this case, a newly established channel with the LURK server will result in a different \texttt{server\_random}.

\subhead{Issues}
Various documents describe two variants of three RSA/DHE modes of LURK for TLS 1.2, plus optional checks (proof of handshake/ownership) with an unclear threat model. Also, discrepancies exist between all documents (e.g., the RFC defines the freshness function as a one-way hash function), prohibiting a clear definition or LURK.
Finally, LURK for TLS 1.3 has not received the same scrutiny.

\subsection{Server-vetted middlebox (renewable content delegation)}
Several approaches achieve content delegation without systematically involving the origin server by relying on special certificates with some degree of revocation, or on binding the server and CDN certificates. The delegation authorization is given for some duration and is renewable. 

\subsubsection{Name constraints}
\label{sec:name-constraints}
The name constraint mechanism~\cite{rfc5280} restricts a CA to issue end entity certificates (EEC) for only a subset of names, e.g., subdomains only.
These constraints can empower an enterprise to possess a CA certificate and issue EEC for its subdomains without the involvement of the CA.
This mechanism could also be used by an end entity (becoming a CA) to issue new browser-trusted certificates for its domains only, which could be shared with CDNs. The proof of delegation becomes apparent from the chain of trust, and the end entity could revoke delegated certificates as needed. Importantly, CDNs do not learn of the EEC private key.

\subhead{Issues}
Due to poor business incentives from CAs, and insufficient support from browsers leading to wrongly accepted certificates,\footnote{Name constraints are not properly enforced by all browsers~\cite{https-meets-cdn}, though the situation is improving, see \url{https://bettertls.com/}} name constraints CA certificates are uncommon.
The entity issuing the EEC also needs to operate as a CA (e.g., revocation servers), although with limited responsibilities (its domains only).

\subsubsection{Proxy certificates}
\label{sec:proxy-certs}
Proxy certificates have been introduced in the context of Grid computing\footnote{Globus Toolkit. \url{https://web.archive.org/web/20070812233445/http://www.globus.org/toolkit/docs/4.0/security/}} to identify Grid users with a PKI. Their purpose was to enable the creation of dynamic identities (e.g., running jobs) and the delegation of some privileges to them (e.g., ability to access user resources, or spawn new jobs).

RFC3820~\cite{rfc3820,proxy-cert04} standardizes Grid proxy certificates as X.509 certificates.
This standard extends the X.509 PKI by allowing EEC to sign proxy certificates (PC) with their own key pair.
PCs could be requested to the end entity by a second entity, making the end entity as a CA.
PCs are also marked with a critical X.509 extension called \textsf{ProxyCertInfo}.
This extension holds a proxy policy, which specifies whether the PC inherits all or none of the privileges of the EEC, or whether an application-specific policy is present to define its privileges. A PC can issue another PC, subject to a separate PC path length constraint. In addition, certain restrictions apply:
1) the Digital Signature key usage bit is required on the issuing EEC;
2) the PC is chained to its issuing EEC (or another PC) by matching the PC issuer to the EEC (or PC) subject, and SANs are forbidden;
3) the PC subject name should be unique and start with the PC issuer name (e.g., it can have an additional common name).
RFC3820 proxy certificates are supported by OpenSSL since v0.9.8~\cite{openssl-changelog}.

Chuat et al.~\cite{delegation-sok} propose a different semantic for proxy certificates, more adapted to delegation on the web. They borrow the idea behind the name constraint mechanism.
Their proposed PCs work with regular non-CA EEC, alleviating the economic and poor support concerns on name constraints, and are constrained to the EEC SANs by default.
The main use case for these PCs would be for CDNs to submit CSRs to the end entity for signature. CDNs could manage their own keys, and the origin server's private key is never shared.

\subhead{Issues}
The lack of SAN support in RFC3820 is prohibitive for modern web usages. Also, the revocation of certificate, and thus of the delegation intent, can only be done by pointing to the end entity as a CRL distribution point or OCSP server, which is impractical since the end entity is not a full-fledged CA.
Chuat et al.'s proposal solve both problems. Notably, revocation is addressed by enforcing short-lived certificates that reduce the attack window after key compromise and enable a fine-grained control over the delegation period. However, it is not supported by clients out-of-the-box.

\subsubsection{Delegated credentials}
\label{sec:delegated-creds}
A relatively new Internet draft~\cite{draft-ietf-tls-subcerts-09} proposes a variant of the proxy certificates defined in RFC3820, by only allowing an end entity with a special EEC to sign \emph{delegated credentials} composed of a public key share and expiration time, bound to be used with TLS 1.3.
The EEC is required to carry a special non-critical extension to permit the use of delegated credentials.
When a client advertises support for this mechanism as part of the TLS extensions, the actual delegated credential is delivered as an extension in the \textsf{CertificateEntry} corresponding to the EEC in the \textsf{Certificate} message, and should be used in place of the EEC public key in the TLS connection (i.e., to verify the \textsf{CertificateVerify} message).
This simplified approach avoids ambiguous interpretations of the semantics of X.509 proxy certificates.

\subhead{Issues} As for proxy certificates, delegated credentials are not supported by clients out-of-the-box.

\subsubsection{Delegation through DANE}
\label{sec:cert-over-dane}
Liang et al.~\cite{https-meets-cdn} propose to leverage DANE~\cite{rfc7671} to associate the server certificate with the CDN certificate used in the actual TLS connection.
DANE is originally intended as a complimentary trust model for TLS that provides constraints on the website certificate in a TLSA DNS record.
In this proposal, the server places its original certificate and that of the CDN itself in TLSA record, thereby asserting the delegation.
When a client connects to the website through the CDN, it receives the CDN certificate valid for the CDN domain name only. The client then retrieves and validates the server certificate obtained through DANE and accepts the CDN certificate for the connection. DANE guarantees that the delegation is approved by the domain owner.

\subhead{Issues}
The support for DNSSEC is still not widespread and has been shown to be prone to misconfiguration~\cite{ChungR0CLMMW17}. In addition, CDNs often employ several certificates for a given domain, requiring constant updates of the TLSA records by the content provider.

\section{Handshaking with middleboxes}
\label{sec:universal-protocols}
When simple TLS splitting and key sharing solutions are not sustainable, in particular due to newly introduced security vulnerabilities~\cite{xcc-tls-proxy,https-intercept-impact} and incompatibilities with TLS 1.3, another approach consists in including middleboxes in the TLS handshake, and assigning them different permissions or selectively disclosing parts of the traffic to them.

\subsection{Selected disclosures}
\label{sec:efgh}
End-to-end Fine Grained HTTP (EFGH~\cite{efgh}) enables policy-based disclosure of encrypted content to a middlebox, and authenticated changes. During the TLS handshake, four symmetric keys are derived: one encryption key $G$ derived from MS known to all parties, another one $K_{CS}$ known to the endpoints only, and two endpoint-to-middlebox keys $K_{CP}$ and $K_{PS}$ for authentication.
The application data is encrypted with $G$ when it is desirable that the content be visible to the proxy, otherwise the end-to-end key $K_{CS}$ is used. $K_{CS}$ also serves to authenticate the records in both cases. When the proxy modifies the traffic, e.g., to deliver content from its cache on behalf of the server, it encrypts the records with $G$ and authenticates them using $K_{CP}$ or $K_{PS}$ depending on the endpoint.

\subhead{Issues}
The protocol requires changes in the TLS record format to accommodate extra headers and metadata, indicating e.g., whether the record is intended to be decrypted by the proxy. In turn, this leads to changes in the client and server implementations. Also, EFGH can only accommodate one middlebox.

\subsection{Permission contexts}
\label{sec:mctls}
Multi-Context TLS (mcTLS~\cite{mctls}) proposes to establish various contexts protected by different sets of symmetric encryption and MAC keys. Each context is managed at the application layer, and corresponds to different middlebox access privileges. Each endpoint performs a key exchange with each middlebox, optionally authenticated. The middleboxes are given half of the context keys by each endpoint for each context they are included in, thus requiring the agreement of both the client and server to access a context. There are two levels of privileges: reader, which receives the reader encryption and MAC keys (for verification only); and writer, which is also given the writer MAC keys (for content modification).

The protocol adds the \textsf{MiddleboxList} extension to the \textsf{ClientHello}, which lists the middleboxes to be included, the labeled contexts and their middlebox permissions. Following the response from the server, each middlebox also responds to each endpoint with their certificate and endpoint-specific key shares (in the case of DHE). In turn, the client is able to derive a distinct shared secret for each middlebox and the endpoint.

The endpoints then derive the endpoint keys that authenticate the traffic to verify that no modification was performed.
They also derive half the reader and writer keys, and send them in \textsf{MiddleboxKeyMaterial} \TLSMsg{MKM} messages to the appropriate middleboxes, encrypted with their corresponding shared secrets. The middleboxes reconstruct the context keys.  An optional feature allows the server to let the client send the whole context keys to the middleboxes.  Writers are able to detect when readers illegally modify the content, while readers can only tell if modifications came from an external attacker. The actual context traffic is encrypted using the reader encryption key; however, endpoints also add MACs to TLS records calculated with the three MAC keys.

The contexts can then be used at the application/server discretion. For instance, HTTP requests and responses could be split into different contexts to enable compression middleboxes to act on the server's response only. Similarly, headers and bodies could be split to enable tracker blockers to perform URL-based filtering without access to the request body and page content. Finally, browsers could decide to enable compression proxies for media content when using metered or slow connections, then dynamically revert to a no-middlebox context when conditions are more favorable.

The authors leave out the provisioning and authentication of authorized middleboxes, and subsequent user interface issues, but point to possible deployment options.

\subhead{Protocol flaws}
Bhargavan et al.~\cite{BhargavanBDFO18} describe two flaws in mcTLS, originating from the lack of formal verification of the protocol. Both attacks assume an attacker who can intercept the traffic on both sides of a middlebox.

The first flaw exploits the absence of a full handshake between the middleboxes and the endpoints, i.e., there are no \textsf{Finish} messages for middleboxes to verify that messages have not been tampered. This allows an attacker who can tamper and revert the \textsf{ClientHello} and \textsf{ServerHello} as they pass through a middlebox to confuse the said middlebox. One attack consists in changing the server name in the SNI extension and returning the corresponding certificate, to lure the middlebox into applying different filtering rules. If the middlebox is additionally caching content, it will misattribute the server response to the wrong server, leading to cache poisoning.

The second flaw exploits the ability of reader middleboxes to modify the content and calculate a valid MAC (as they are given the reader MAC key), which can confuse other readers down the path before being rejected by the endpoints.
Those flaws consider a powerful attacker with the ability to intercept network communications simultaneously at different locations between the endpoints, which may not be realistic in certain cases.

\subheadit{TLMSP}
Transport layer Middlebox Security Protocol~\cite{tlmsp} is a recently standardized protocol based on mcTLS aiming to bring multiple contexts with the ability to let middleboxes read and insert or delete information. Furthermore, it brings on-path middlebox discovery where mcTLS suggested to pre-configure the clients, making it more deployment-friendly. The standard offers a way for middleboxes to add themselves in a list in the \textsf{ClientHello}, which is echoed back to the client in the \textsf{ServerHello} for the client to accept them. Similarly, a server could add server-side middleboxes. Also, TLMSP fixes the previously reported flaws on mcTLS by introducing a new \textsf{Finished} message (\textsf{MboxFinished}) to conclude the handshake between the endpoints and each middlebox.
Other improvements include a guarantee that the messages went through all middleboxes and in the intended order. This property is also guaranteed in maTLS (see §\ref{sec:matls}). 
However, TLMSP appears to add complexity compared to mcTLS due to e.g., newly introduced features such as the middlebox discovery process that leads to a negotiation possibly spanning across multiple TCP connections, the ability for a middlebox to leave an ongoing session, and session resumption. A careful verification of the protocol is missing.

\subsection{Publicly auditable middleboxes}
\label{sec:matls}
Unlike mcTLS, Middlebox-Aware TLS (maTLS~\cite{matls}) proposes to negotiate separate TLS sessions between middleboxes instead of sharing the same session secrets.
However, contrary to simply splitting the sessions (§\ref{sec:split-sessions}), maTLS enables both endpoints to supervise each segment's negotiated parameters, and for the client to authenticate both the middeboxes and the end-server.
This architecture also enables the endpoints to detect which specific middlebox modified the traffic, even when two of them have write access.

During the \textsf{ClientHello}/\textsf{ServerHello} exchange, the participating middleboxes append their certificate to the server \textsf{Certificate} message, and their DH key share. The client is able to validate each certificate. Each endpoint and middlebox establishes a TLS session with the next entity, under which the traffic will be encrypted. The middleboxes derive pairwise accountability keys $ak_{i,j}$ with both endpoints, which are used to verify the security parameters $p_{i,j}$ negotiated between each segment.
When forwarding or modifying a message, a middlebox appends a Modification Log (ML) to the message, which consists of a chain of HMACs on the message received and the one sent by each middlebox. The client is able to verify which middlebox modified the message, and that the message went through all the middleboxes in the right order.

Middlebox service providers (MSP) should obtain a valid publicly trusted certificate with critical extensions to assert the requested access level (read or write) to endpoints. Middlebox certificates are logged into Middlebox Transparency (MT) logs, similar to CT logs, to provide auditability of issuance.

\subhead{Issues}
The authors claim that maTLS ``allows middleboxes to participate in the TLS session in a visible and auditable fashion.''
However, maTLS does not achieve auditability of \emph{participation in a TLS session}, which could be interpreted as the ability to detect illegitimate traffic inspection after-the-fact.
Rather, an audit of MT logs only reveals which MSP has been issued a certificate (which could be viewed as a ``license to intercept'').
Consequently, while each middlebox is implicitly vetted by the CA that issued its special-purpose certificate, the end user is still involved in the decision process to accept those certificates.
Given that middlebox certificates cannot be bound to a unique and unambiguous identity, the end user is vulnerable to phishing-like attacks whereby an attacker obtains a certificate with a convincing name, or even the same name as the expected MSPs by justifying the name in another jurisdiction (cf.~\cite{ev-stripe}).
Paradoxically, by enforcing reliable identification of the providers, small enterprise middleboxes and every instance of client-end software solutions (e.g., antivirus) would be prevented from obtaining such certificates, making maTLS either as a non-usable or incomplete protocol.

\section{Privacy-preserving inspection}
\label{sec:privacy-inspection}
A different approach to traffic inspection consists in limiting the exposure of data to middleboxes by providing evidence to the endpoints that the operations performed are trustworthy instead of blindly trusting middleboxes once they gain read and or write capabilities.

\subsection{Searchable encryption-based solutions}
\label{sec:blindbox}
Techniques based on searchable encryption have been proposed to preserve the privacy of filtered traffic.

BlindBox~\cite{blindbox} enables for middleboxes to perform basic keyword searching and pattern matching on TLS traffic without revealing all of the data to the middleboxes.
The authors consider two privacy models that enable different types of operations. In the exact match privacy model, only keyword matching is possible while not revealing the rest of the data. In the probable cause privacy model, the traffic can be decrypted for further processing (e.g., regular expression matching) if a keyword is found.
The filtering rules are provided by a trusted third party. BlindBox leverages garbled circuits and oblivious transfer so that the middlebox does not learn the session key and the rules are not known by the client and server.
It requires each sender to tokenize, encrypt and send their traffic separately from TLS, for the middlebox to compare the tokens against the rules.

\subhead{Issues}
At least one receiver should be honest and detect whether the TLS and BlindBox traffic differ. More importantly, BlindBox incurs a 500\% bandwidth overhead, and requires 97 seconds to perform the initial TLS handshake for a typical IDS solution with 3,000 rules.
The use cases presented by the authors include the detection of malware installed on student laptops connected to the university network, and the filtering of traffic by an ISP/enterprise in accordance to the user's preference (e.g., parental control, document watermarking). We note that in the former case, the malware needs to comply with the BlindBox protocol, which is unlikely. Indeed, both the malware and the remote server may be controlled by the attacker, defeating BlindBox's threat model. In the latter case, BlindBox needs to be widely deployed on web servers for the user to truly benefit from the filters.

The performance of BlindBox is significantly improved in PrivDPI~\cite{privdpi} by replacing expensive garbled circuits with modular exponentiations. However, it was later shown that PrivDPI middleboxes could in some cases recover intermediate reusable computations and adapt them to generate any rule, enabling middleboxes to perform arbitrary keyword matching to recover the plaintext~\cite{NingHPXLWD20}. The authors of this attack also propose a more efficient rule generation that does not require the middlebox to precompute rules with the client and server.

\subsection{Using Trusted Execution Environments}
\label{sec:sgx-box}
Rather than involving cryptographic solutions, other proposals rely on the guarantees of Trusted Execution Environments (TEEs) such as Intel Software Guard Extensions (SGX) with remote attestation, to protect long-term keys or even the whole middlebox operations while operating on the traffic itself with low overhead. 

SGX-Box~\cite{sgx-box} proposes to run middlebox capabilities in SGX enclaves. It relies on memory isolation to keep the computations private, and on remote attestation to guarantee that the middlebox program executed is the intended one.
The origin server establishes an out-of-band long-term channel with the on-path middlebox and performs a remote attestation of its enclave before sharing TLS session secrets for each client connection.
Clients could also request an attestation; however, it is unclear how end users or enterprises enforce the use of specific middleboxes as the server is responsible for sharing TLS secrets.
Moreover, SGX-Box introduces a high-level language that abstracts packet parsing and TLS decryption operations to perform middlebox operations.

STYX~\cite{styx} improves on the performance of Keyless SSL~\cite{keyless-ssl} by provisioning the origin's private key and keeping private key operations in SGX enclaves at the CDN. However, it does not restrict the usage of the private key, which could turn the enclave into an oracle on a compromised edge.
Phoenix~\cite{phoenix} resembles STYX but is designed to port existing applications to SGX enclaves while also restricting the use of the private key by attested applications only. Presumably, the origin server is able to attest the permitted software stack before provisioning the private key with an edge server. Furthermore, the CDN can be made agnostic to the cached content thanks to memory and file encryption managed in enclaves.

mbTLS~\cite{mbtls} enables endpoints to communicate per-segment session keys to middleboxes after the software running in enclaves is remotely attested, possibly fitting various use cases.
However, note that SGX-provided guarantees have been eroded by numerous attacks, see~\cite{foreshadow,sgxpectre,vanbulck2020lvi} and especially against remote attestation~\cite{sgaxe}. Although microcode and software updates address these attacks, the technology seems particularly affected by side-channel attacks.

\section{Evaluation}
\label{sec:evaluation}
In this section, we evaluate the consistency of the techniques and proposals discussed in §\ref{sec:enterprise-twists},\ref{sec:cdn},\ref{sec:universal-protocols},\ref{sec:privacy-inspection} against the actual use cases we identified in §\ref{sec:challenges-tls} for access to HTTP traffic in the presence of TLS. We consider the compatibility of these schemes with participating stakeholders' incentives evaluated in~§\ref{sec:stakeholders} (in particular in Table~\ref{tab:stakes}).
Second, we set out an evaluation framework aimed at characterizing the deployability of those schemes along with their security features and guarantees, and run each scheme through it.

\begin{table*}[t!]
 \centering
 \footnotesize
 \setlength{\tabcolsep}{1.5pt}
 \caption{Stakeholder alignment and conceivable applicability of approaches to enable content access according to middlebox use cases.
 We consider the end-user organization imposes its incentives on end users and clients in use cases \textit{d}, \textit{f}, \textit{h}-CN and \textit{j}-CN.
 Legend: Circle portioning qualifies the middlebox use case applicability. \yes=applicable, \hmm=partially applicable, \maybe=potentially applicable but is not proposed as such (proposals only), \no=not applicable.  Mismatching incentives lead to a gray circle (\killedByI{\yes}, \killedByI{\hmm}, \killedByI{\maybe}).
 CD=Client device, CN=Client network, SN=Content provider (server) network.
 }
 \begin{threeparttable}
  \resizebox{\textwidth}{!}{%
  \begin{tabular}
   {
    c|
    c|
    p{36mm}|
    c|
    c||
    c|
    c|
    c|
    c|
    c|
    c|
    c|
    c|
    c|
    c|
    c|
    c|
    c|
    c||
    c|
    c|
    c|
    c|
    c|
    c|
    c|
    c|
    c||
    c|
    c|
    c|
    c|
    c|
    c|
   }
   \multicolumn{1}{c|}{} &
   \rotatebox{90}{\bfseries{Discussed in §}} &
   \diagbox[width=37mm]{\bfseries{Scheme name}}{\rotatebox{0}{\bfseries{Use cases}}} &
   \mcrot{1}{l|}{90}{\emph{a,b,c}. ISP legal operations} &
   \mcrot{1}{l|}{90}{\itm{d} Fraud detection} &
   \mcrot{2}{l|}{90}{\itm{e} Malware download prevention} &
   \mcrot{1}{l|}{90}{\itm{f1} IDS on outgoing connection} &
   \mcrot{1}{l|}{90}{\itm{f2} IPS on outgoing connection} &
   \mcrot{1}{l|}{90}{\itm{g1} IDS on incoming connection} &
   \mcrot{2}{l|}{90}{\itm{g2} IPS/WAF on incoming conn.} &
   \mcrot{2}{l|}{90}{\itm{h} Data retention (compliance)} &
   \mcrot{1}{l|}{90}{\itm{i} Pre-request authorization} &
   \mcrot{2}{l|}{90}{\itm{j} Data loss prevention} &
   \mcrot{1}{l|}{90}{\itm{k1} IoT dev.\ monitoring (server auth.)} &
   \mcrot{1}{l|}{90}{\itm{k2} IoT dev.\ monitoring (mutual auth.)} &
   \mcrot{3}{l|}{90}{\itm{l} L7 load-balancing} &
   \mcrot{1}{l|}{90}{\itm{m} Broadcasting} &
   \mcrot{1}{l|}{90}{\itm{n} Server-mandated caching} &
   \mcrot{2}{l|}{90}{\itm{o} Opportunistic caching} &
   \mcrot{2}{l|}{90}{\itm{p} Transcoding/compression} &
   \mcrot{1}{l|}{90}{\itm{q1} Problem troubleshooting (user help)} &
   \mcrot{1}{l|}{90}{\itm{q2} Problem troubleshooting (no user help)} &
   \mcrot{3}{l|}{90}{\itm{r} Parental control} &
   \mcrot{1}{l|}{90}{\itm{s} ISP billing/tracking} \\
   
   \cline{3-34}
   \multicolumn{1}{c|}{~} &
   \multicolumn{1}{c|}{~} &
   \multicolumn{1}{r|}{\bfseries{Middlebox location $\rightarrow$}} &
   {\scriptsize ISP} &
   {\scriptsize CN} &
   
   {\scriptsize CD} &
   {\scriptsize CN} &
   
   {\scriptsize CN} &
   {\scriptsize CN} &
   
   {\scriptsize SN}  &
   {\scriptsize SN}  &
   {\scriptsize CDN} &
  
   {\scriptsize CN} &
   {\scriptsize SN} &
   
   {\scriptsize CDN} &
   
   {\scriptsize CN} &
   {\scriptsize SN} &
   
   \multicolumn{2}{c|}{\scriptsize CN} &
   
   {\scriptsize ISP} &
   {\scriptsize CDN} &
   {\scriptsize SN} &
   
   {\scriptsize ISP} &
   {\scriptsize CDN} &
   
   {\scriptsize CN} &
   {\scriptsize ISP} &
   
   {\scriptsize ISP} &
   {\scriptsize CDN} &
   
   \multicolumn{2}{c|}{\scriptsize SN} &
   
   {\scriptsize CD} &
   {\scriptsize CN} &
   {\scriptsize ISP} &
   
   {\scriptsize ISP} \\
   
   \noalign{\hrule height 1pt}
   \multirow{12.5}{*}{\rotatebox{90}{\centering Session splitting/key sharing}} &
   \multirow{3}{*}{\ref{sec:split-sessions}} &
   TLS session splitting &
    \killedByI{\yes} &
    \yes &
    \yes &
    \yes &
    \hmm &
    \yes &
    \killedByI{\hmm} &
    \killedByI{\yes} &
    \killedByI{\yes} &
    \yes &
    \killedByI{\yes} &
    \killedByI{\yes} &
    \yes &
    \killedByI{\yes} &
    \killedByI{\yes} &
    \no &
    \killedByI{\yes} &
    \killedByI{\yes} &
    \killedByI{\yes} &
    \no &
    \killedByI{\yes} &
    \yes &
    \killedByI{\yes} &
    \killedByI{\yes} &
    \killedByI{\yes} &
    \yes &
    \killedByI{\yes} &
    \yes &
    \yes &
    \killedByI{\yes} &
    \killedByI{\yes} \\
   ~&
   ~&
   \hspace{1em}w/ ProxyInfoExtension~\cite{draft-mcgrew-tls-proxy-server-01} &
    \killedByI{\yes} &
    \yes &
    \yes &
    \yes &
    \hmm &
    \yes &
    \killedByI{\hmm} &
    \killedByI{\yes} &
    \killedByI{\yes} &
    \yes &
    \killedByI{\yes} &
    \killedByI{\yes} &
    \yes &
    \killedByI{\yes} &
    \killedByI{\yes} &
    \no &
    \killedByI{\yes} &
    \killedByI{\yes} &
    \killedByI{\yes} &
    \no &
    \killedByI{\yes} &
    \yes &
    \killedByI{\yes} &
    \killedByI{\yes} &
    \killedByI{\yes} &
    \yes &
    \killedByI{\yes} &
    \yes &
    \yes &
    \killedByI{\yes} &
    \killedByI{\yes} \\
   ~&
   ~&
   \hspace{1em}w/ client-trusted CA &
    \yes &
    {\yes} &
    {\yes} &
    {\yes} &
    {\hmm} &
    {\yes} &
    \hmm &
    \yes &
    {\yes} &
    {\yes} &
    \yes &
    {\yes} &
    {\yes} &
    \yes &
    \yes &
    \no &
    {\yes} &
    {\yes} &
    \yes &
    \no &
    {\yes} &
    {\yes} &
    {\yes} &
    {\yes} &
    {\yes} &
    \yes &
    \yes &
    {\yes} &
    {\yes} &
    {\yes} &
    {\yes} \\
   ~&
   \ref{sec:share-priv-key} &
   \hspace{1em}w/ server certificate &
    \killedByI{\yes} &
    \killedByI{\yes} &
    \killedByI{\yes} &
    \killedByI{\yes} &
    \killedByI{\hmm} &
    \killedByI{\yes} &
    \hmm &
    \yes &
    {\yes} &
    \killedByI{\yes} &
    \yes &
    {\yes} &
    \killedByI{\yes} &
    \yes &
    \yes &
    \no &
    \killedByI{\yes} &
    {\yes} &
    \yes &
    \no &
    {\yes} &
    {\yes} &
    \killedByI{\yes} &
    \killedByI{\yes} &
    {\yes} &
    \yes &
    \yes &
    {\yes} &
    {\yes} &
    \killedByI{\yes} &
    \killedByI{\yes} \\
   ~&
   \ref{sec:share-priv-key} &
   Certificate sharing (passive) &
    \killedByI{\yes} &
    \killedByI{\yes} &
    \killedByI{\yes} &
    \killedByI{\yes} &
    \killedByI{\yes} &
    \killedByI{\yes} &
    \yes &
    \yes &
    \yes &
    \killedByI{\yes} &
    \yes &
    \no &
    \killedByI{\yes} &
    \yes &
    \killedByI{\yes} &
    \killedByI{\yes} &
    \no & 
    \no & 
    \no & 
    \no &
    \no &
    \no &
    \no &
    \no &
    \no &
    \yes &
    \yes &
    \killedByI{\yes} &
    \killedByI{\yes} &
    \killedByI{\yes} &
    \killedByI{\yes} 
   \\
    
   ~&
   \ref{sec:dh-key-share-vanilla} &
   Static DH key sharing (server) &
    \killedByI{\yes} &
    \killedByI{\yes} &
    \killedByI{\yes} &
    \killedByI{\yes} &
    \killedByI{\yes} &
    \killedByI{\yes} &
    \yes &
    \yes &
    \yes &
    \killedByI{\yes} &
    \yes &
    \no &
    \killedByI{\yes} &
    \yes &
    \killedByI{\yes} &
    \killedByI{\yes} &
    \no &
    \no &
    \no &
    \no &
    \no &
    \no &
    \no &
    \no &
    \no &
    \yes &
    \yes &
    \killedByI{\yes} &
    \killedByI{\yes} &
    \killedByI{\yes} &
    \killedByI{\yes} 
    
   \\
   \cline{2-34}

   ~&
   \multirow{2}{*}{\ref{sec:ets}} &
   ETS~\cite{ets} (scenario 1\&3) &
    \killedByI{\maybe} &
    \yes &
    \no & 
    \no &
    \yes &
    \no &
    \no &
    \no &
    \no &
    \yes &
    \no &
    \no &
    \yes &
    \no &
    \killedByI{\maybe} & 
    \no & 
    \no &
    \no &
    \no &
    \no &
    \no &
    \no &
    \no &
    \no &
    \no &
    \no &
    \no &
    \no &
    \no &
    \maybe &
    \killedByI{\maybe} 
    
   \\
   ~&~&
   ETS~\cite{ets} (scenario 2) &
    \no &
    \no &
    \no &
    \no &
    \no &
    \no &
    \yes &
    \no &
    \no &
    \no &
    \yes &
    \no &
    \no &
    \yes &
    \no & 
    \no & 
    \no &
    \no &
    \no &
    \no &
    \no &
    \no &
    \no &
    \no &
    \no &
    \yes &
    \yes &
    \no &
    \maybe &
    \no &
    \no 
    
   \\
   ~&
   \ref{sec:tls-visibility} &
   tls\_visibility~\cite{draft-rhrd-tls-tls13-visibility-01} &
    \killedByI{\maybe} &
    \no &
    \no &
    \no &
    \no &
    \no &
    \killedByI{\yes} &
    \killedByI{\yes} &
    \killedByI{\maybe} &
    \no &
    \killedByI{\yes} &
    \no &
    \no &
    \killedByI{\yes} &
    \killedByI{\maybe} &
    \killedByI{\maybe} & 
    \no &
    \no &
    \no &
    \no &
    \killedByI{\maybe} &
    \no &
    \no &
    \no &
    \killedByI{\maybe} &
    \yes &
    \killedByI{\yes} &
    \no &
    \no &
    \no &
    \no
   \\
   ~&
   \ref{sec:sslkeylog} &
   SSLKEYLOGFILE~\cite{sslkeylog-nss} &
    \no &
    \hmm &
    \yes &
    \no &
    \no &
    \no &
    \no &
    \no &
    \no &
    \no &
    \no &
    \no &
    \hmm &
    \no &
    \no &
    \no &
    \no &
    \no &
    \no &
    \no &
    \no &
    \no &
    \no &
    \no &
    \no &
    \no &
    \no &
    \yes &
    \no &
    \no &
    \no
    
   \\
   ~&
   \ref{sec:sslkeylog} &
   LOCKS~\cite{locks} &
    \killedByI{\maybe} &
    \yes &
    \yes &
    \yes &
    \yes &
    \yes &
    \killedByI{\maybe} &
    \killedByI{\maybe} &
    \killedByI{\maybe} &
    \yes &
    \killedByI{\maybe} &
    \no &
    \yes &
    \killedByI{\maybe} &
    \killedByI{\yes} &
    \killedByI{\yes} &
    \no &
    \no &
    \no &
    \no &
    \killedByI{\maybe} &
    \maybe &
    \killedByI{\maybe} &
    \killedByI{\maybe} &
    \killedByI{\maybe} &
    \yes &
    \killedByI{\yes} &
    \yes &
    \yes &
    \yes &
    \killedByI{\yes}
    
   \\
   ~&
   \ref{sec:tls-rar} &
   TLS-RaR~\cite{tls-rar} &
    \killedByI{\maybe} &
    \maybe &
    \no &
    \no &
    \hmm &
    \no &
    \killedByI{\hmm} &
    \no &
    \no &
    \maybe &
    \killedByI{\maybe} &
    \no &
    \no &
    \no &
    \killedByI{\yes} &
    \killedByI{\yes} &
    \no &
    \no &
    \no &
    \no &
    \no &
    \no &
    \no &
    \no &
    \no &
    \maybe &
    \killedByI{\maybe} &
    \no &
    \no &
    \no &
    \no 
    
   \\
    
   \hline
   
   \multirow{7.5}{*}{\rotatebox{90}{\centering Delegation}} &
   \ref{sec:untrusted-caches} &
   Barnraising~\cite{Lesniewski-LaasK05} &
    \killedByI{\maybe} &
    \no &
    \killedByI{\maybe} &
    \killedByI{\maybe} &
    \no &
    \no &
    \no &
    \no &
    \no &
    \no &
    \no &
    \no &
    \no &
    \no &
    \no &
    \no &
    \no &
    \no &
    \no &
    \no &
    \yes &
    \killedByI{\yes} &
    \killedByI{\yes} &
    \no &
    \no &
    \no &
    \no &
    \killedByI{\maybe} &
    \killedByI{\maybe} &
    \killedByI{\maybe} &
    \no
   \\
   
   ~&
   \ref{sec:untrusted-caches} &
   Akamai's patent~\cite{gero2015splicing} &
   
    \killedByI{\maybe} &
    \no &
    \no &
    \no &
    \no &
    \no &
    \maybe &
    \maybe &
    {\maybe} &
    \no &
    \maybe &
    \no &
    \no &
    \maybe &
    \killedByI{\maybe} &
    \killedByI{\maybe} &
    \no &
    \no &
    \no &
    \no &
    {\yes} &
    \no &
    \no &
    \no &
    {\maybe} &
    \maybe &
    \maybe &
    \no &
    \no &
    \no &
    \no
     
   \\
   ~& \ref{sec:qos23} &
   QoS2~\cite{qos2}, QoS3~\cite{al2019qos3} &

    \no &
    \no &
    \no &
    \no &
    \no &
    \no &
    \no &
    \no &
    \no &
    \no &
    \no &
    \no &
    \no &
    \no & 
    \no &
    \no &
    \no &
    \no &
    \no &
    \no &
    \no &
    \killedByI{\yes} &
    \killedByI{\yes} &
    \no &
    \no &
    \no &
    \no &
    \no &
    \no &
    \no &
    \no

\\

   ~&
   \ref{sec:server-authd-caching} &
   Keyless SSL~\cite{keyless-ssl}, LURK~\cite{BoureanuMPAMFM20} &
    \killedByI{\maybe} &
    \killedByI{\maybe} &
    \killedByI{\maybe} &
    \killedByI{\maybe} &
    \killedByI{\hmm} &
    \killedByI{\maybe} &
    \maybe &
    \maybe &
    {\maybe} &
    \killedByI{\maybe} &
    \maybe &
    \no &
    \killedByI{\maybe} &
    \maybe &
    \killedByI{\maybe} &
    \killedByI{\maybe} &
    \killedByI{\maybe} &
    \maybe &
    \maybe &
    \no &
    {\yes} &
    \killedByI{\maybe} &
    \killedByI{\maybe} &
    \killedByI{\maybe} &
    {\maybe} &
    \maybe &
    \maybe &
    \killedByI{\maybe} &
    \killedByI{\maybe} &
    \killedByI{\maybe} &
    \killedByI{\maybe} 

    \\
  
	~&
	\ref{sec:name-constraints} &
	Name constraints~\cite{rfc5280}&
	\killedByI{\yes} &
	\killedByI{\yes} &
	\killedByI{\yes} &
	\killedByI{\yes} &
	\killedByI{\hmm} &
	\killedByI{\yes} &
	\hmm &
	\yes &
	\yes &
	\killedByI{\yes} &
	\yes &
	\yes &
	\killedByI{\yes} &
	\killedByI{\yes} &
	\killedByI{\yes} &
	\no &
	\killedByI{\yes} &
	\yes &
	\yes &
	\no &
	\yes &
	\killedByI{\yes} &
	\killedByI{\yes} &
	\killedByI{\yes} &
	\yes &
	\yes &
	\yes &
	\killedByI{\yes} &
	\killedByI{\yes} &
	\killedByI{\yes} &
	\killedByI{\yes}
	
	\\
   ~&
   \ref{sec:proxy-certs} &
   Proxy certificates~\cite{delegation-sok} &
    \killedByI{\yes} &
    \killedByI{\yes} &
    \killedByI{\yes} &
    \killedByI{\yes} &
    \killedByI{\hmm} &
    \killedByI{\yes} &
    \killedByI{\hmm} &
    \killedByI{\yes} &
    \killedByI{\yes} &
    \killedByI{\yes} &
    \killedByI{\yes} &
    \killedByI{\yes} &
    \killedByI{\yes} &
    \killedByI{\yes} &
    \killedByI{\yes} &
    \no &
    \killedByI{\yes} &
    \killedByI{\yes} &
    \killedByI{\yes} &
    \no &
    \killedByI{\yes} &
    \killedByI{\yes} &
    \killedByI{\yes} &
    \killedByI{\yes} &
    \killedByI{\yes} &
    \yes &
    \killedByI{\yes} &
    \killedByI{\yes} &
    \killedByI{\yes} &
    \killedByI{\yes} &
    \killedByI{\yes}
    
   \\
 
   ~&
   \ref{sec:delegated-creds} &
   Delegated credentials~\cite{draft-ietf-tls-subcerts-09} &
   
    \killedByI{\yes} &
    \killedByI{\yes} &
    \killedByI{\yes} &
    \killedByI{\yes} &
    \killedByI{\hmm} &
    \killedByI{\yes} &
    \killedByI{\hmm} &
    \killedByI{\yes} &
    \killedByI{\yes} &
    \killedByI{\yes} &
    \killedByI{\yes} &
    \killedByI{\yes} &
    \killedByI{\yes} &
    \killedByI{\yes} &
    \killedByI{\yes} &
    \no &
    \killedByI{\yes} &
    \killedByI{\yes} &
    \killedByI{\yes} &
    \no &
    \killedByI{\yes} &
    \killedByI{\yes} &
    \killedByI{\yes} &
    \killedByI{\yes} &
    \killedByI{\yes} &
    \yes &
    \killedByI{\yes} &
    \killedByI{\yes} &
    \killedByI{\yes} &
    \killedByI{\yes} &
    \killedByI{\yes}
    
   \\
   ~&
   \ref{sec:cert-over-dane} &
   Delegation over DANE~\cite{https-meets-cdn} &
  
    \killedByI{\yes} &
    \killedByI{\yes} &
    \killedByI{\yes} &
    \killedByI{\yes} &
    \killedByI{\hmm} &
    \killedByI{\yes} &
    \killedByI{\hmm} &
    \killedByI{\yes} &
    \killedByI{\yes} &
    \killedByI{\yes} &
    \killedByI{\yes} &
    \killedByI{\yes} &
    \killedByI{\yes} &
    \killedByI{\yes} &
    \killedByI{\yes} &
    \no &
    \killedByI{\yes} &
    \killedByI{\yes} &
    \killedByI{\yes} &
    \no &
    \killedByI{\yes} &
    \killedByI{\yes} &
    \killedByI{\yes} &
    \killedByI{\yes} &
    \killedByI{\yes} &
    \yes &
    \killedByI{\yes} &
    \killedByI{\yes} &
    \killedByI{\yes} &
    \killedByI{\yes} &
    \killedByI{\yes}
    
   \\
   
   \hline
   \multirow{4.5}{*}{\rotatebox{90}{\centering 3-way}} &
   \ref{sec:efgh} &
   EFGH~\cite{efgh} &
    \killedByI{\yes} &
    \killedByI{\yes} &
    \killedByI{\yes} &
    \killedByI{\yes} &
    \killedByI{\hmm} &
    \killedByI{\yes} &
    \killedByI{\hmm} &
    \killedByI{\yes} &
    \killedByI{\yes} &
    \killedByI{\yes} &
    \killedByI{\yes} &
    \no &
    \killedByI{\yes} &
    \killedByI{\yes} &
    \killedByI{\yes} &
    \killedByI{\yes} &
    \no &
    \no &
    \no &
    \no &
    \killedByI{\yes} &
    \killedByI{\maybe} &
    \killedByI{\maybe} &
    \killedByI{\yes} &
    \killedByI{\yes} &
    \yes &
    \killedByI{\yes} &
    \killedByI{\yes} &
    \killedByI{\yes} &
    \killedByI{\yes} &
    \killedByI{\yes} 
    
   \\
   ~&
   \multirow{1}{*}{\ref{sec:mctls}} &
   mcTLS~\cite{mctls} &
    \killedByI{\yes} &
    \killedByI{\yes} &
    \killedByI{\yes} &
    \killedByI{\yes} &
    \killedByI{\hmm} &
    \killedByI{\yes} &
    \killedByI{\hmm} &
    \killedByI{\yes} &
    \killedByI{\yes} &
    \killedByI{\yes} &
    \killedByI{\yes} &
    \no &
    \killedByI{\yes} &
    \killedByI{\yes} &
    \killedByI{\yes} &
    \killedByI{\yes} &
    \no & 
    \no & 
    \no & 
    \no &
    \killedByI{\yes} &
    \killedByI{\yes} &
    \killedByI{\yes} &
    \killedByI{\yes} &
    \killedByI{\yes} &
    \yes &
    \killedByI{\yes} &
    \killedByI{\yes} &
    \killedByI{\yes} &
    \killedByI{\yes} &
    \killedByI{\yes} 
   
   \\
   ~&
   \multirow{1}{*}{\ref{sec:mctls}} &
   TLMSP~\cite{tlmsp} &
    \killedByI{\yes} &
    \killedByI{\yes} &
    \killedByI{\yes} &
    \killedByI{\yes} &
    \killedByI{\hmm} &
    \killedByI{\yes} &
    \killedByI{\hmm} &
    \killedByI{\yes} &
    \killedByI{\yes} &
    \killedByI{\yes} &
    \killedByI{\yes} &
    \no &
    \killedByI{\yes} &
    \killedByI{\yes} &
    \killedByI{\yes} &
    \killedByI{\yes} &
    \no & 
    \no & 
    \no & 
    \no &
    \killedByI{\yes} &
    \killedByI{\yes} &
    \killedByI{\yes} &
    \killedByI{\yes} &
    \killedByI{\yes} &
    \yes &
    \killedByI{\yes} &
    \killedByI{\yes} &
    \killedByI{\yes} &
    \killedByI{\yes} &
    \killedByI{\yes}

   \\
   
   ~&
   \ref{sec:matls} &
   maTLS~\cite{matls} &
    \killedByI{\yes} &
    \yes &
    \yes &
    \yes &
    \hmm &
    \yes &
    \killedByI{\hmm} &
    \killedByI{\yes} &
    \killedByI{\yes} &
    \yes &
    \killedByI{\yes} &
    \killedByI{\maybe} &
    \yes &
    \killedByI{\yes} &
    \killedByI{\yes} &
    \killedByI{\yes} &
    \no & 
    \no & 
    \no & 
    \no &
    \killedByI{\yes} &
    \killedByI{\yes} &
    \killedByI{\yes} &
    \killedByI{\yes} &
    \killedByI{\yes} &
    {\yes} &
    \killedByI{\yes} &
    \yes &
    \yes &
    \yes &
    \killedByI{\yes}

   \\

   \hline 
   
   \multirow{3.5}{*}{\rotatebox{90}{\centering Others}} &
   \ref{sec:blindbox} &
   BlindBox~\cite{blindbox}, PrivDPI~\cite{privdpi} &
    \no &
    \killedByI{\maybe} &
    \killedByI{\yes} &
    \killedByI{\yes} &
    \killedByI{\yes} &
    \killedByI{\yes} &
    \killedByI{\maybe} &
    \killedByI{\maybe} &
    \killedByI{\maybe} &
    \killedByI{\yes} &
    \killedByI{\maybe} &
    \no &
    \killedByI{\maybe} &
    \killedByI{\maybe} &
    \no & 
    \no &
    \no &
    \no &
    \no &
    \no &
    \no &
    \no &
    \no &
    \no &
    \no &
    \no &
    \no &
    \killedByI{\maybe} &
    \killedByI{\maybe} &
    \killedByI{\maybe} &
    \no 
   \\

   \cline{2-34}
   ~&
   \ref{sec:sgx-box} &
   SGX-Box, mbTLS~\cite{sgx-box,mbtls} &
    \killedByI{\maybe} &
    \no &
    \no &
    \no &
    \no &
    \no &
    \yes &
    \yes &
    {\yes} &
    \no &
    \maybe &
    \no &
    \no &
    \no &
    \killedByI{\maybe} &
    \killedByI{\maybe} &
    \no &
    \no &
    \no &
    \no &
    {\maybe} &
    \no &
    \no &
    \no &
    {\maybe} &
    \maybe &
    \maybe &
    \no &
    \no &
    \no &
    \no
    \\

   ~&
   \ref{sec:sgx-box} &
   STYX, Phoenix~\cite{styx,phoenix} &
    \killedByI{\maybe} &
    \no &
    \no &
    \no &
    \no &
    \no &
    \yes &
    \yes &
    {\yes} &
    \no &
    \maybe &
    \yes &
    \no &
    \no &
    \killedByI{\maybe} &
    \killedByI{\maybe} &
    \no &
    \no &
    \no &
    \no &
    {\maybe} &
    \no &
    \no &
    \no &
    {\maybe} &
    \maybe &
    \maybe &
    \no &
    \no &
    \no &
    \no
    \\
   
   \hline
   
  \end{tabular}}\par
  
 \end{threeparttable}
 \label{tab:proposal-usage}
\vspace{-.2in}
\end{table*}

\subsection{Disincentives for use cases}
We evaluate in Table~\ref{tab:proposal-usage} how techniques and proposals match with the stakeholders' incentives for each use case of a middlebox, i.e., whether schemes only require support from stakeholders that have clear motivating incentives (capital letters in Table~\ref{tab:stakes}).
This evaluation aims to identify proposals that are ill-suited for specific use cases due to obvious misalignment of incentives. In particular, it does not account for technical deployability concerns and security features, which are further evaluated in §\ref{sec:evaluation-framework}. The criteria for applicability only considers whether the scheme enables the required access to unencrypted content for the use case. For instance, it is conceivable that proxy certificates are applicable for the \emph{IDS on outgoing connections (f1)} use case, despite the content provider's obvious reluctance to provide such certificates to the EUO in the general case.
For general techniques such as session splitting and certificate sharing, we consider the applicability against all use cases, despite foolish combinations, e.g., session splitting at the server-side that requires provisioning the client with a custom root certificate (a situation typically seen at the client-side). This serves as a comparison for other proposals.
From this simple evaluation, we are able to make the following observations.

\subhead{Lack of incentives invalidates several proposals}
From Table~\ref{tab:proposal-usage}, it is apparent that many seemingly applicable scheme-use case pairs become ill-suited due to a lack of incentives from one of the main stakeholders. Irrespective of further security and deployment evaluations, this analysis shows that some proposals, although promising on other aspects, are likely stopped from even being considered due to their design.
In particular, schemes that require the collaboration of both endpoints for all applicable use cases are necessarily deployment-unfriendly. Such schemes could be beneficial if the sum of the use cases they solve may convince all involved stakeholders to support them.
For instance, BlindBox is targeted at filtering traffic using third-party rules selected by the client; however, it requires the server's active participation.
Conversely, \textsf{tls\_visibility} cannot achieve any use case beyond problem troubleshooting with the user's help since it mandates the client's participation for server-side middlebox operations. The same applies to delegation through special certificates (e.g., proxy certificates, delegated credentials); however, see further discussion in §\ref{sec:findings}.

\subhead{No universal solutions}
EFGH, mcTLS and maTLS are presented as universal solutions to accommodate any sort of common middlebox~\cite{efgh,mctls,matls}. However, EFGH and mcTLS require the support of the server and all three require support of the client as well, which is incompatible with the incentives of the end user, their organization, or the content provider. For instance, to support malware download prevention, mcTLS requires the support from the server, which is mostly uninterested in this task.
As a consequence from the previous point,
these general-purpose solutions do not achieve the intended goal of giving more control over the presence of middleboxes; rather, they force an unrelated endpoint to support its solution with no clear benefit. They would only be useful if they replaced TLS. Arguably, such a move deserves strong justifications.
Of interest, maTLS benefits from a per-segment construction (a ``bottom-up'' approach as the authors describe), which enables use cases where middleboxes are operated near the client-side. Although the client does not fully benefit from the protocol's features when the server does not support it, middleboxes can be accommodated.

\subhead{No convincing scheme for ISP legal operations}
All schemes that could in theory apply to ISP legal use cases are obviously incompatible with the end user's incentives as the client needs to support the scheme (e.g., session splitting with non-browser trusted certificate), and with the server operator's incentives if required to participate (e.g., mcTLS). Although TLS session splitting w/ client-trusted CA achieves the technical need, it is incompatible with CA certificate issuance policies.

\subhead{Unaddressed use cases}
Consumer IoT device monitoring suffers from a lack of incentive from the main stakeholder, i.e., the vendor, and therefore no scheme seems fit for this purpose. It seems necessary that a business incentive (e.g., through certifications) or legal requirements be created for IoT device vendors to open interfaces that would let end users monitor the traffic originating from these devices.
Also, broadcasting and L7 load balancing are special use cases for which little to no scheme applies. Indeed, broadcasting involves distributing the same content to many receivers which is incompatible with end-to-end encryption. The latter use case (especially for mobile ISPs) is probably not sufficiently motivated to attract researchers, while also being subject to performance constraints.

\subsection{Evaluation framework}
\label{sec:evaluation-framework}
We describe here the criteria that enable the evaluation of the proposals and techniques discussed in the previous sections. Those criteria derive from our incentive analysis, important issues and limitations we have highlighted for each scheme, and desirable properties aimed by each scheme.
As various objectives require different criteria, we focus our evaluation on three main categories of use cases: CDN-type (N) middleboxes (\emph{g}, \emph{i}, \emph{l}, \emph{n} and \emph{p}), client-side (C) and server-side (R) middleboxes. Our evaluation is based on general and case-specific criteria for deployability, security, usability and extra features provided. Our ratings are illustrated in Table~\ref{tab:proposal-eval-cdn}.

\subsubsection{Deployability}
Schemes that can be readily deployed while being compatible with existing up-to-date major browsers fulfill \emph{Client Compatibility (D1)}, while those that require minor modifications to the clients (typically supporting an additional TLS extension), partially fulfill it. Conversely, \emph{Server Compatibility (D2)} is fulfilled when schemes are compatible with existing servers. We do not expect schemes to fulfill both criteria, and decide which one to evaluate depending on the use case.
Also, schemes that do not require any changes in the application running on top of TLS (client or server-side) fulfill \emph{No Application-layer Rewrite (D3)}. If changes are only needed to better take advantage of the scheme, they partially fulfill the criteria.
When schemes enable the client with technical means of knowing about the presence of middleboxes owned or operated by a third party fulfill \emph{Client Awareness (D4)}. A full mark is given if the client can fully distinguish all middleboxes; half mark if the client may only know if any number of middleboxes is present or not.
Some schemes may rely on features that are not widely supported, such as DNSSEC on the client, while others fulfill \emph{No Unsupported Prerequisite (D5)}. Requiring SGX on the middlebox partially fulfills the criteria.
Schemes where the establishment of a secure connection by the client does not require additional connections to exchange side information with middleboxes or the other endpoint fulfill \emph{No Add.\ Connections (D6)}. When an endpoint and middlebox maintain a permanent connection used for all TLS connections, they partially fulfill D6.
Schemes that preserve the existing PKI and do not involve CAs and CT logs in extra activities fulfill \emph{No Extra Burden on PKI (D7)}.
If schemes only require new or unusual X.509 extensions or parameters in TLS certificates issued by trusted CAs, they partially fulfill the criteria.
When the scheme does not introduce extra servers or new roles for existing servers, it fulfills \emph{No Suppl.\ Infrastructure (D8)}.
Schemes that only introduce a reasonable network bandwidth and computation overhead fulfill \emph{Reasonable Overhead (D9)}. Here, reasonable means that the overhead does not appear to be prohibitive for deployment and user experience.

\subsubsection{Security and privacy}
Simple schemes leverage the knowledge of the server's certificate private key or other long-term keys.
When those secrets are not exposed to third party-owned middleboxes, schemes fulfill \emph{Long-term Secret Stays Private (S1)}.
Ciphersuites recommended in TLS 1.2 and all those in TLS 1.3 comprise AEAD symmetric ciphers. When schemes can encrypt the request and content served by the end-server and middleboxes with these ciphers, they fulfill \emph{Support for AEAD (S2)}.
Similarly, recommended key exchanges in TLS 1.2 and all those in TLS 1.3 are based on Diffie-Hellman. When schemes truly leverage ephemeral keys in all key exchanges from which derived keys are used to encrypt the request and content, they fulfill \emph{Support for FS (S3)}.
Schemes that allow the client to verify and eventually choose the ciphersuite (CS) selections on all segments between both ends fulfill \emph{Client CS Supervision (S4)}.
Schemes that enable the client to see and validate the end-server certificate fulfill \emph{End-server Authentication (S5)}.
The connection should be cryptographically bound to this certificate to receive a full mark, i.e., encryption secrets must be derived from a key exchange authenticated (directly or not) by that certificate's private key. Schemes where the middlebox is trusted to expose the end-server certificate receive a half mark.
If there is a way for the most interested end party to verify or limit operations done by a middlebox on the traffic to ensure that, e.g., it does not retain or process data in an unexpected way, the schemes fulfill \emph{Operation Verifiability (S6)}. This can be provided by cryptographic means, or through TEEs.

\subhead{Sub-case: CDN}
In the event of an edge server compromise, schemes that prevent the attacker to break past recorded communications made through this or other edges fulfill \emph{FS Despite Edge Compromise (N-SA)}. If the prevention relies on negotiating relevant FS ciphersuites among non-FS ones, as with TLS 1.2, they partially fulfill N-SA. We assume TLS 1.2 is used as the base protocol when not constrained by the schemes. Full prevention can also be achieved if schemes fulfill S6.
Schemes that enable the client to validate the end-server certificate (S5), and verify that the CDN delegation is approved by the content provider fulfill \emph{Origin \& CDN Authentication (N-SB)}.
The connection should be cryptographically bound to both verifications.
When the origin server can control and restrict all content served by the CDN, schemes fulfill \emph{Origin-controlled Delivery (N-SC)}. Note that N-SC is implied from S6.

\subhead{Sub-case: Client/server-side solutions}
The content provider is the only one to decide of its own architecture, including middleboxes. When the client cannot opt out of those middleboxes, schemes fulfill \emph{No Bypass of Server-side MB (R-SA)}.
When the traffic is modified by a client-side middlebox, schemes fulfill \emph{Modification Accountability (C-SA)} if the modification is easily detectable by mechanisms provided in the protocol (e.g., verification of a cryptographic hash), and the middlebox responsible for the modification is unambiguously identifiable (provided all middleboxes are first authenticated). Also, a middlebox that is only allowed to read the traffic cannot modify it without being detected in a similar fashion.
If the identification of the middleboxes only return the set of middleboxes that have the ability to modify the traffic, the proposal receives a half-mark.

\subsubsection{Usability and other options}
Ideally, the end user should not need to make new security-related decisions, due to the inherent risk that users make the wrong choice. Schemes that do not involve the user into making new decisions related to middleboxes fulfill \emph{No User Involvement (U1)}.
When schemes can help to solve other strong problems in the TLS ecosystem, they fulfill \emph{Extra Benefits (O1)}. For instance, proxy certificates and delegated credentials enable short-lived certificates/credentials to be issued without the involvement of CAs, thereby alleviating the need for (problematic) revocation mechanisms.
Schemes that are compatible with TLS client authentication fulfill \emph{Support for TLS Client Auth (O2)}.

\subhead{Sub-case: Client/server-side solutions}
Permanent client-side middleboxes can be configured on end-user systems to be trusted, in which case schemes fulfill \emph{Pre-configured Trust (C-OA)}. When unknown middleboxes can be discovered by the schemes, they fulfill \emph{Middlebox Discovery (C-OB)}.
If a scheme can accommodate multiple middleboxes in the connection, it fulfills \emph{Multi-MB Support (C-OC)}.
The scheme fulfills \emph{Private Mode (C-OD)} if the client can opt out of inspection for selected connections.

\begin{table*}[t]
 \centering
 \footnotesize
 \setlength{\tabcolsep}{1.5pt}
 \caption{Comparative evaluation of schemes considering three use cases: N for CD\underline{N}, R for se\underline{r}ver-side, C for \underline{c}lient-side security middleboxes, * means all three. Notation for entries: \yes=property fulfilled by a scheme; \hmm=partially fulfilled; \no=not fulfilled. Shaded regions mean property not applicable to specified use case.}
 \hspace{-.4in}\begin{threeparttable}
  \begin{tabular}
   {
    c|
    p{36mm}|
    c||
    c
    c
    c
    c
    c
    c
    c
    c
    c|
    c
    c
    c
    c
    c
    c|
    c
    c
    c
    c
    c|
    c|
    c
    c|
    c
    c
    c
    c|
   }
 \rotatebox{90}{\bfseries{Discussed in §}} & \diagbox[width=37mm]{\bfseries{Scheme name}}{\rotatebox{0}{\bfseries{Criteria}}} & \mcrot{1}{l}{60}{\bfseries{Use Case}} & \mcrot{1}{l}{60}{\bfseries{D1. Client Compatibility}} & \mcrot{1}{l}{60}{\bfseries{D2. Server Compatibility}} & \mcrot{1}{l}{60}{\bfseries{D3. No Application-layer Rewrite}} & \mcrot{1}{l}{60}{\bfseries{D4. Client Awareness}} & \mcrot{1}{l}{60}{\bfseries{D5. No Unsupported Prerequisite}} & \mcrot{1}{l}{60}{\bfseries{D6. No Add.\ Connections}}& \mcrot{1}{l}{60}{\bfseries{D7. No Extra Burden on PKI}}& \mcrot{1}{l}{60}{\bfseries{D8. No Suppl.\ Infrastructure}}& \mcrot{1}{l}{60}{\bfseries{D9. Reasonable Overhead}}& \mcrot{1}{l}{60}{\bfseries{S1. Long-term Secret Stays Private}} & \mcrot{1}{l}{60}{\bfseries{S2. Support for AEAD}} & \mcrot{1}{l}{60}{\bfseries{S3. Support for FS}} & \mcrot{1}{l}{60}{\bfseries{S4. Client CS Supervision}} & \mcrot{1}{l}{60}{\bfseries{S5. End-server Authentication}} & \mcrot{1}{l}{60}{\bfseries{S6. Operation Verifiability}} & \mcrot{1}{l}{60}{\bfseries{N-SA. FS Despite Edge Compromise}} & \mcrot{1}{l}{60}{\bfseries{N-SB. Origin \&  CDN Authentication}} & \mcrot{1}{l}{60}{\bfseries{N-SC. Origin-controlled Delivery}} & \mcrot{1}{l}{60}{\bfseries{C-SA. Modification Accountability}} & \mcrot{1}{l}{60}{\bfseries{R-SA. No Bypass of Server-side MB}} & \mcrot{1}{l}{60}{\bfseries{U1. No Risky Decisions}}& \mcrot{1}{l}{60}{\bfseries{O1: Extra Benefits}} & \mcrot{1}{l}{60}{\bfseries{O2: Support for TLS Client Auth}} & \mcrot{1}{l}{60}{\bfseries{C-OA. Pre-configured Trust}} & \mcrot{1}{l}{60}{\bfseries{C-OB. Middlebox Discovery}} & \mcrot{1}{l}{60}{\bfseries{C-OC. Multi-MB Support}}& \mcrot{1}{l}{60}{\bfseries{C-OD. Private Mode}}

\\

\noalign{\hrule height 1pt}
\multirow{3}{*}{\ref{sec:split-sessions}} & TLS session splitting &*& \no &\yes& \yes & \yes & \yes &\yes&\yes&\yes&\yes& \yes & \yes &\yes&\no&\no& \no & \hmm & \no & \no &\no&\yes&\yes&\no&\no&\yes&\no&\hmm&\hmm
\\

~& \hspace{1em}w/ ProxyInfoExtension~\cite{draft-mcgrew-tls-proxy-server-01} &*& \no &\yes& \yes & \yes & \yes &\yes&\yes&\yes&\yes& \yes & \yes &\yes&\no&\hmm& \no & \hmm & \hmm & \no &\no&\yes&\yes&\no&\no&\yes&\no&\hmm&\hmm

\\

~& \hspace{1em}w/ client-trusted CA &*& \yes & \yes& \yes & \no & \yes &\yes&\hmm&\yes&\yes& \no & \yes &\yes&\no&\no& \no & \hmm & \no & \no &\no&\yes&\yes&\no&\no&\no&\no&\yes&\hmm

\\

\ref{sec:share-priv-key} & \hspace{1em}w/ server certificate &*& \yes &\yes& \yes & \no & \yes &\yes&\yes&\yes&\yes& \no & \yes &\yes&\no&\hmm& \no & \hmm & \no & \no &\no&\yes&\yes&\no&\no&\no&\no&\yes&\hmm

\\

\ref{sec:share-priv-key}&Certificate sharing (passive)&R&\yes&\yes&\yes&\no&\yes&\yes&\yes&\yes&\yes&\no&\yes&\no&\yes&\yes&\no&\cellcolor{gray!15}&\cellcolor{gray!15}&\cellcolor{gray!15}&\cellcolor{gray!15}&\yes&\yes&\no&\yes&\cellcolor{gray!15}&\cellcolor{gray!15}&\cellcolor{gray!15}&\cellcolor{gray!15}

\\

\ref{sec:dh-key-share-vanilla}&Static DH key sharing (server)&R&\yes&\hmm&\yes&\hmm&\yes&\yes&\yes&\yes&\yes&\no&\yes&\no&\yes&\yes&\no&\cellcolor{gray!15}&\cellcolor{gray!15}&\cellcolor{gray!15}&\cellcolor{gray!15}&\yes&\yes&\no&\yes&\cellcolor{gray!15}&\cellcolor{gray!15}&\cellcolor{gray!15}&\cellcolor{gray!15}

\\
\hline

\ref{sec:ets}&ETS~\cite{ets} (scenario 1)&R&\hmm&\no&\yes&\hmm&\yes&\yes&\hmm&\no&\yes&\hmm&\yes&\no&\yes&\yes&\no&\cellcolor{gray!15}&\cellcolor{gray!15}&\cellcolor{gray!15}&\cellcolor{gray!15}&\yes&\yes&\no&\yes&\cellcolor{gray!15}&\cellcolor{gray!15}&\cellcolor{gray!15}&\cellcolor{gray!15}

\\
&ETS~\cite{ets} (scenario 2)&R&\yes&\no&\yes&\no&\yes&\yes&\hmm&\no&\yes&\no&\yes&\no&\yes&\yes&\no&\cellcolor{gray!15}&\cellcolor{gray!15}&\cellcolor{gray!15}&\cellcolor{gray!15}&\yes&\yes&\no&\yes&\cellcolor{gray!15}&\cellcolor{gray!15}&\cellcolor{gray!15}&\cellcolor{gray!15}
\\

&ETS~\cite{ets} (scenario 3)&R&\no&\yes&\yes&\yes&\yes&\yes&\yes&\no&\yes&\yes&\yes&\no&\no&\yes&\no&\cellcolor{gray!15}&\cellcolor{gray!15}&\cellcolor{gray!15}&\cellcolor{gray!15}&\yes&\yes&\no&\yes&\cellcolor{gray!15}&\cellcolor{gray!15}&\cellcolor{gray!15}&\cellcolor{gray!15}

\\

\ref{sec:tls-visibility}&tls\_visibility~\cite{draft-rhrd-tls-tls13-visibility-01}&N,R&\no&\cellcolor{gray!15}&\yes&\yes&\yes&\yes&\yes&\yes&\yes&\yes&\yes&\yes&\yes&\yes&\no&\yes&\no&\no&\cellcolor{gray!15}&\no&\yes&\no&\yes&\cellcolor{gray!15}&\cellcolor{gray!15}&\cellcolor{gray!15}&\cellcolor{gray!15}

\\

\ref{sec:sslkeylog}&SSLKEYLOGFILE~\cite{sslkeylog-nss}&C&\no&\yes&\yes&\yes&\yes&\yes&\yes&\yes&\yes&\yes&\yes&\yes&\yes&\yes&\no&\cellcolor{gray!15}&\cellcolor{gray!15}&\cellcolor{gray!15}&\cellcolor{gray!15}&\cellcolor{gray!15}&\yes&\no&\yes&\yes&\no&\hmm&\no

\\

\ref{sec:sslkeylog}&LOCKS~\cite{locks}&C&\no&\yes&\yes&\yes&\yes&\hmm&\yes&\no&\yes&\yes&\yes&\yes&\yes&\yes&\no&\cellcolor{gray!15}&\cellcolor{gray!15}&\cellcolor{gray!15}&\cellcolor{gray!15}&\cellcolor{gray!15}&\yes&\no&\yes&\yes&\no&\hmm&\no

\\

\ref{sec:tls-rar}&TLS-RaR~\cite{tls-rar}&C&\no&\yes&\yes&\yes&\yes&\yes&\yes&\yes&\yes&\yes&\yes&\yes&\yes&\yes&\no&\cellcolor{gray!15}&\cellcolor{gray!15}&\cellcolor{gray!15}&\cellcolor{gray!15}&\cellcolor{gray!15}&\yes&\no&\yes&\yes&\no&\hmm&\no

\\
\hline

\ref{sec:untrusted-caches}&Barnraising~\cite{Lesniewski-LaasK05}&N&\yes&\cellcolor{gray!15}&\yes&\no&\yes&\yes&\yes&\no&\yes&\yes&\no&\yes&\yes&\yes&\no&\yes&\no&\yes&\cellcolor{gray!15}&\cellcolor{gray!15}&\yes&\no&\yes&\cellcolor{gray!15}&\cellcolor{gray!15}&\cellcolor{gray!15}&\cellcolor{gray!15}

\\

\ref{sec:enc-mac-splitting} & Akamai's patent~\cite{gero2015splicing} &N& \yes &\cellcolor{gray!15}& \yes & \no & \yes &\yes&\yes&\yes&\yes& \yes & \no &\no&\yes&\yes& \no & \yes & \no & \yes &\cellcolor{gray!15}&\cellcolor{gray!15}&\yes&\no&\yes&\cellcolor{gray!15}&\cellcolor{gray!15}&\cellcolor{gray!15}&\cellcolor{gray!15}

\\

\ref{sec:qos23}&QoS2~\cite{qos2}, QoS3~\cite{al2019qos3}&N&\no&\cellcolor{gray!15}&\no&\yes&\yes&\no&\hmm&\yes&\yes&\yes&\yes&\yes&\no&\no&\no&\yes&\no&\yes&\cellcolor{gray!15}&\cellcolor{gray!15}&\yes&\no&\no&\cellcolor{gray!15}&\cellcolor{gray!15}&\cellcolor{gray!15}&\cellcolor{gray!15}

\\

\ref{sec:keyless-ssl} & Keyless SSL~\cite{keyless-ssl} &N,R& \yes &\cellcolor{gray!15}& \yes & \no & \yes &\no&\yes&\no&\yes& \yes & \yes &\yes&\no&\yes& \no & \no & \no &\no&\cellcolor{gray!15}&\yes&\yes&\no&\no&\cellcolor{gray!15}&\cellcolor{gray!15}&\cellcolor{gray!15}&\cellcolor{gray!15}

\\

\ref{sec:lurk} & LURK~\cite{BoureanuMPAMFM20} &N,R& \yes &\cellcolor{gray!15}& \yes & \no & \yes &\no&\yes&\no&\yes& \yes & \yes &\yes&\no&\yes& \no & \yes & \no &\no&\cellcolor{gray!15}&\yes&\yes&\no&\no&\cellcolor{gray!15}&\cellcolor{gray!15}&\cellcolor{gray!15}&\cellcolor{gray!15}

\\

\ref{sec:name-constraints} & Name constraints~\cite{rfc5280}&*& \hmm &\yes& \yes & \yes & \hmm &\yes&\hmm&\yes&\yes& \yes & \yes &\yes&\no&\yes& \no & \hmm & \yes & \no &\no&\yes&\yes&\yes&\yes&\no&\no&\yes&\hmm
\\

\ref{sec:proxy-certs} & Proxy certificates~\cite{delegation-sok} &*& \no &\yes& \yes & \yes & \yes &\yes&\yes&\yes&\yes& \yes & \yes &\yes&\no&\yes& \no & \hmm & \yes & \no &\no&\yes&\yes&\yes&\yes&\no&\no&\yes&\hmm
\\

\ref{sec:delegated-creds} & Delegated credentials~\cite{draft-ietf-tls-subcerts-09} &*& \hmm &\yes& \yes & \yes & \yes &\yes&\hmm&\yes&\yes& \yes & \yes &\yes&\no&\yes& \no & \yes & \yes & \no &\no&\yes&\yes&\yes&\yes&\no&\no&\hmm&\yes

\\

\ref{sec:cert-over-dane} & Delegation over DANE~\cite{https-meets-cdn} &*& \no &\yes& \yes & \yes & \no &\yes&\no&\yes&\yes& \yes & \yes &\yes&\no&\yes& \no & \hmm & \yes & \no &\no&\yes&\yes&\no&\yes&\no&\no&\hmm&\hmm

\\
\hline

\ref{sec:efgh}&EFGH~\cite{efgh}&*&\no&\no&\hmm&\yes&\yes&\yes&\yes&\yes&\yes&\yes&\yes&\yes&\yes&\yes&\no&\yes&\yes&\no&\yes&\no&\no&\no&\yes&\yes&\no&\no&\yes

\\

\multirow{1}{*}{\ref{sec:mctls}} & mcTLS~\cite{mctls} &*& \no &\no& \hmm & \yes & \yes &\yes&\yes&\yes&\hmm& \yes & \yes &\yes&\yes&\yes& \no & \hmm & \yes & \hmm &\hmm&\no&\no&\no&\yes&\yes&\no&\yes&\yes

\\

\multirow{1}{*}{\ref{sec:mctls}} & TLMSP~\cite{tlmsp} &*& \no &\no& \hmm & \yes & \yes &\yes&\yes&\yes&\hmm& \yes & \yes &\yes&\yes&\yes& \no & \hmm & \yes & \hmm &\hmm&\no&\no&\no&\yes&\yes&\yes&\yes&\yes
\\

\ref{sec:matls} & maTLS~\cite{matls} &*& \no &\hmm& \yes & \yes & \yes &\yes&\no&\yes&\yes& \yes & \yes &\yes&\yes&\yes& \no & \hmm & \yes & \hmm &\yes&\no&\no&\no&\yes&\yes&\yes&\yes&\yes
\\
\hline

\ref{sec:blindbox} & BlindBox~\cite{blindbox} &R& \no &\no & \yes & \yes & \yes &\no&\yes&\no&\no& \yes & \yes &\yes&\yes&\yes& \yes & \cellcolor{gray!15}&\cellcolor{gray!15}&\cellcolor{gray!15}& \cellcolor{gray!15}&  \no&\yes&\no&\yes&\cellcolor{gray!15}&\cellcolor{gray!15}&\cellcolor{gray!15}&\cellcolor{gray!15}

\\

\ref{sec:blindbox} & PrivDPI~\cite{privdpi} &R& \no &\no & \yes & \yes & \yes &\no&\yes&\no&\hmm& \yes & \yes &\yes&\yes&\yes& \yes & \cellcolor{gray!15}&\cellcolor{gray!15}&\cellcolor{gray!15}& \cellcolor{gray!15}&  \no&\yes&\no&\yes&\cellcolor{gray!15}&\cellcolor{gray!15}&\cellcolor{gray!15}&\cellcolor{gray!15}

\\

\ref{sec:sgx-box} & SGX-Box~\cite{sgx-box} &N& \yes &\cellcolor{gray!15}& \yes & \no & \hmm &\no&\yes&\no&\yes& \yes & \yes &\yes&\yes&\yes& \yes & \yes & \no & \yes &\cellcolor{gray!15}&\cellcolor{gray!15}&\yes&\no&\yes&\cellcolor{gray!15}&\cellcolor{gray!15}&\cellcolor{gray!15}&\cellcolor{gray!15}

\\

\ref{sec:sgx-box} & STYX~\cite{styx} &N&\yes&\cellcolor{gray!15}& \yes &\no&\hmm&\hmm&\yes&\no&\yes&\hmm&\yes&\yes&\yes&\yes&\no&\hmm&\no&\no&\cellcolor{gray!15}&\cellcolor{gray!15}&\yes&\no&\yes&\cellcolor{gray!15}&\cellcolor{gray!15}&\cellcolor{gray!15}&\cellcolor{gray!15}
\\

\ref{sec:sgx-box} & mbTLS~\cite{mbtls} (server-side MB)&N,R&\yes&\cellcolor{gray!15}& \yes &\no&\hmm&\yes&\yes&\yes&\yes&\yes&\yes&\yes&\yes&\yes&\yes&\yes&\no&\yes&\cellcolor{gray!15}&\yes&\yes&\no&\yes&\cellcolor{gray!15}&\cellcolor{gray!15}&\cellcolor{gray!15}&\cellcolor{gray!15}
\\
\ref{sec:sgx-box} & mbTLS~\cite{mbtls} (client-side MB)&C&\no&\yes& \yes &\no&\hmm&\yes&\yes&\yes&\yes&\yes&\yes&\yes&\yes&\yes&\yes&\cellcolor{gray!15}&\cellcolor{gray!15}&\cellcolor{gray!15}&\no&\cellcolor{gray!15}&\yes&\no&\yes&\yes& \yes&\yes&\yes
\\

\ref{sec:sgx-box} & Phoenix~\cite{phoenix} &N&\yes&\cellcolor{gray!15}& \yes &\no&\hmm&\hmm&\yes&\yes&\yes&\hmm&\yes&\yes&\yes&\yes&\yes&\hmm&\no&\yes&\cellcolor{gray!15}&\cellcolor{gray!15}&\yes&\no&\yes&\cellcolor{gray!15}&\cellcolor{gray!15}&\cellcolor{gray!15}&\cellcolor{gray!15}
\\
   \hline
  \end{tabular}
 \end{threeparttable}
 \label{tab:proposal-eval-cdn}
 \vspace{-.1in}
\end{table*}

\section{Discussion and concluding remarks}
\label{sec:discussion}
We list below important findings and discuss open questions that arise from our analysis.

\subsection{Findings to highlight}
\label{sec:findings}
\subhead{Generic solutions do not solve any use case well} Table~\ref{tab:proposal-usage} shows the misalignment of stakeholders incentives that affects in particular generic solutions such as EFGH, mcTLS, and maTLS.
These generic solutions require support from both ends and are therefore less likely to be deployed. Moreover, they lack essential properties in specific use cases. For instance, server-side middleboxes are completely bypassable if the client decides not to support the protocol (R-SA in Table~\ref{tab:proposal-eval-cdn}). Unless these new proposals are universally deployed, users could easily opt to use regular TLS instead.
A better option seems to develop solutions tailored for specific use cases and that are aligned with stakeholder goals.

\subhead{Lack of well-defined goals}
From our reading of the proposed protocols, different research proposals seem to take many directions, sometimes without precise objectives, and often address different problems, which complicates comparative evaluation.
Without clearly specified goals and evaluations, it is difficult to appreciate whether a proposed scheme meets the objectives.
Our discussion of practical use cases for access to plaintext (§\ref{sec:challenges-tls}) is intended to address this issue.

\subhead{Two main approaches for CDNs}
The CDN use case is approached in two different ways in the literature: either give more autonomy to the middlebox without releasing the origin's long-term secret, or control the operations of the middlebox through a TEE to only release origin secrets to a secure enclave (or hide them through other cryptographic means). These two approaches highlight an underlying difference in trust assumptions. Which of these two approaches is better would appear to depend on specific contexts.

\subhead{Performance-enhancing proxies cannot be reliably discovered}
The discovery or automated provisioning of client-side middleboxes (up to the ISP included), e.g., for network-dependent optimization purposes, is a challenging task that is often overlooked or considered orthogonal in the schemes we surveyed. For example, while mcTLS authors suggest provisioning proxy parameters through DHCP (arguably an insecure design), maTLS relies on CAs to issue special and unrestricted publicly trusted middlebox certificates; neither of these approaches seem securely deployable. EFGH is silent regarding provisioning.
More recently, TLMSP allows middleboxes to notify endpoints of their presence as a request to get included. However, proper authentication of those middleboxes is not fully addressed. Other proposals do not address this use case. This leaves us with incomplete solutions.

\subhead{Delegated credentials offer to solve multiple problems} In Nov.\ 2019, the delegated credentials (DC) mechanism was implemented by Cloudflare and Firefox (in nightly builds), with the support from DigiCert~\cite{delegated-creds-cf}.
The early adoption of DC, despite a misalignment of stakeholder incentives in our view (Table~\ref{tab:proposal-usage}), is due to another advantage of this mechanism. In practice, DC brings the equivalent of short-lived certificates, at low cost, which mitigates the certificate revocation issue by reducing the attack window from months or years to days~\cite{delegation-sok}. Proxy certificates bring a similar benefit.
Thus, this may indicate that proposals addressing other open issues may succeed by solving multiple hard problems at the same time.
Interestingly, we note that DC is conceptually simpler compared to academic efforts.

\subhead{TLS 1.3 analysis efforts are not yet matched}
Formal verification of protocols led to the discovery of vulnerabilities in Keyless SSL~\cite{keyless-ssl-analysis} and mcTLS~\cite{BhargavanBDFO18}.
It is unclear whether other proposals have received such peer scrutiny.
Given the complexity of TLS and the formal treatments it received, we expect that any significant modifications to the protocol may introduce weaknesses if not carefully reviewed by the community and verified by e.g., formal methods.
With the numerous proposals in the literature, we hope our study will serve to target effort towards the most promising ones in priority.
On this path, it is worth highlighting the effort of Meier et al.~\cite{MeierSCB13} and Bhargavan et al.~\cite{BhargavanBDFO18} in analyzing TLS proxying procotols using formal verification frameworks.

\subhead{Usability concerns}
For client-side middleboxes that require the client's approval or validation, the end-user interface and experience is poorly studied. For instance, should the URL bar display separate secure locks in the browser for each middlebox? Should the user consent to the middlebox for each connection, or should she be given the choice to enable/disable the middlebox? If a middlebox certificate is expired, can the user bypass the browser warning, and for how long?
TLMSP allows several ambiguous situations to happen that would require the end user to make a decision to allow (or not) newly discovered middleboxes in a connection. For instance, it is unclear how users would treat requests such as: ``\textit{would you allow the transparent middlebox with URI: ThisIsYourAntivirus, IPv4: 192.168.1.12, and the attached self-signed certificate, to modify your traffic to google.com?}''
Such issues further complicate HTTPS usability, and are under-studied.

\subsection{Middlebox privacy policy}
While websites disclose their privacy policy to users, which may relate to e.g., data retention practices, middleboxes might follow different rules. This is particularly relevant when the middleboxes are not owned by either endpoints, e.g., for network performance enhancements. Privacy and legal issues may arise, especially in view of the European General Data Protection Regulation (GDPR)~\cite{gdpr} and related privacy laws. We have not explicitly considered within this paper the privacy and legal issues; these appear to be open questions that further complicate potential deployability. Note that these issues are seldom discussed in the literature and would appear to deserve, we suggest, more attention.

\subsection{Alternatives to interception}
For certain use cases, there are research avenues and industrially deployed solutions that aim at understanding parts of the content hidden with TLS without decryption.
The key idea is that encrypted traffic leaks some information that enables an adversary to understand parts of its content~\cite{traffic-classification}. For instance, the number of TCP packets exchanged, their size and pace, can be used to fingerprint which website or even which page is being visited~\cite{web-fingerprint}.
Defenses against this fingerprinting typically incur significant overhead~\cite{Walkie-Talkie}.
The traffic metadata can also be used to distinguish traffic from benign or malicious applications~\cite{AndersonPM18,cisco-eta}, based on advertised and negotiated ciphersuites, TLS version and extensions.
Anderson et al.~\cite{AndersonCDM19} also showed that specific HTTP headers could be inferred from HTTPS traffic.
Reed et al.~\cite{ReedK17} could identify the videos played on Netflix over HTTPS by analyzing the TCP/IP headers only.
Ede et al.~\cite{FlowPrint} propose a machine learning approach for creating mobile application fingerprints from encrypted network traffic.
Thus, it appears that specific scenarios could be addressed by deployable alternative solutions to interception and delegation.
\subsection{Other open issues}
From our analysis, we highlight the following open questions and issues to consider.
\begin{enumerate}
\item When a TLS connection is terminated at a CDN, load balancer or firewall, and may not even reach the intended end-server, can it be called end-to-end? In some cases, the CDN acts as a host provider and could be considered a legitimate end. Similarly, if an enterprise terminates incoming TLS connections at an edge node (as in ETS), should the user be given control of the traffic up to an application server?

\item Proposals such as ME-TLS~\cite{me-tls} and TwinPeaks~\cite{twinpeaks} rely on identity-based encryption (IBE) to achieve efficient key negotiation between endpoints and middleboxes for IoT monitoring.
While pushing for IBE as a replacement for the web PKI is unrealistic, there might be a use for it in the IoT ecosystem. 
\item TLS inspection may negatively impact protocol performance improvements such as TLS session resumption~\cite{delegation-sok} and 0-RTT. Moreover, performance-related use cases for access to plaintext traffic may not even benefit from traffic inspection proposals due to computational costs. In such a case, the middlebox could be given the information it seeks (e.g., preferred bitrate, type of content) as part of cleartext metadata that accompanies the TLS traffic, as done in EFGH; unfortunately, EFGH itself suffers from deployment challenges.
\item Emerging countries and rural areas often suffer from poor connectivity/speed, low coverage and data quotas, making optimizations important. However, if supporting performance-enhancing proxies require changes in the HTTPS ecosystem, they will likely not be supported by all relevant stakeholders (e.g., content providers). Furthermore, the practice of customizing browsers for satellite-based ISP customers~\cite{lepeska-proxy} is suboptimal.

\end{enumerate}

\subsection{Concluding remarks}
Each common network practice for middleboxes that relies on access to plaintext TLS traffic comes with different requirements and implications.
For the use case for Content Delivery Networks, which includes content caching and DDoS attack prevention, from our analysis it would appear that there are technically mature solutions, albeit not widely deployed, such as delegated credentials, proxy certificates, or trusted hardware-based solutions; for example, Cloudflare is pursuing delegated credentials~\cite{delegated-creds-cf} as discussed.
For most other use cases, despite abundant enthusiasm in the literature, the existence of practical, deployable solutions has yet to be demonstrated; from our analysis, we expect that deployment barriers may be too hard to overcome.
In particular, we have identified important open issues surrounding middlebox-friendly proposals including an increased complexity, expected end-user issues in transparency and understanding, and the lack of support for existing TLS performance optimization mechanisms.
Also, it appears that industry arguments for plaintext access for compliance purposes are not explicitly supported by accompanying evidence (see our analysis of PCI-DSS and NERC CIP in §\ref{sec:industry-reg}).
Importantly, an interesting question that appears not to have been seriously explored is what alternatives to TLS interception exist to address individual use cases, and which better align with everyone's incentives.
Moreover, while there is a typical stakeholder conflict for operations motivated by law enforcement or regulatory controls enforced at the ISP, other conflicts such as the use of performance-enhancing middleboxes in high latency/low speed scenarios (e.g., rural areas or otherwise poorly served in Internet services) highlight difficult tradeoffs.
Ultimately, given the importance of TLS---and it would appear that deployments and major service providers including Google are leaning towards HTTPS being the default replacing HTTP~\cite{chrome-http-not-secure}---existing common network practices that rely on plaintext will apparently stop working or lose functionality.  We encourage the community to explicitly discuss and reach consensus on the requirements for acceptable middlebox solutions, e.g., as an informational IETF RFC or working group draft, in particular considering specific use cases. 
This would ideally trigger discussion between industry and academic experts to agree on common grounds and focus the research effort to address specific issues. For example, one would expect that these requirements would include aspects such as rigorous independent protocol analysis before deployment (as with TLS 1.3), access to plaintext consistent with least privilege principles and privacy considerations, and accountability for any modification of content.
A recent effort by public and private sector enterprises~\cite{ccpl-workshop} helps to document a few of their requirements, e.g., scalability, effectiveness, but appears to prioritize access at the expense of the goals of privacy enthusiasts.
We encourage similar effort but with a broader spectrum of stakeholders.

\section*{Acknowledgments}
The first author acknowledges funding from a DND/NSERC Discovery Grant supplement.
The second author acknowledges funding from the Natural Sciences and Engineering Research Council of Canada (NSERC) for both his Canada Research Chair in Authentication and Computer Security, and a Discovery Grant. We thank the CCSL and CISL members at Carleton University for their valuable feedback.


\appendix

\section{Sample protocol runs for interception schemes}
\label{apx:illustrations}
The 14 illustrations below provide a high-level view of sample protocol runs of several schemes discussed in this survey to highlight their major features and to make it easier to identify key differences. We omitted some details on purpose, but investigated the authors' implementation when available to fill gaps. When several modes, options or versions are available, we prefer to showcase TLS 1.2 with ECDHE key exchange.

\underline{Notation:} C is the client, MB is the middlebox, S is the server. We explicitly list the secrets known to a party in ($\cdot$) when needed for clarity. ($P_\text{S}^+/P_\text{S}^-$) and ($P_\text{M}^+/P_\text{M}^-$) are, resp., the server and middlebox certificate key pairs, $(c, s, m_1, m_2)$ are, resp., the private keys of the client, server, client-side and server-side middlebox. $r_i$ is a random number from entity $i$. See §\ref{sec:background-messages} for background information on the TLS messages.
We also make a number of simplifications for the sake of conciseness.
For instance, when ECDHE public parameters are exchanged, the standard format from RFC8422~\cite{rfc8422} requires a structured description of the elliptic curve followed by a point coordinates. We simply denote by $mG$ these public parameters, with $m$ being a secret (\underline{m}iddlebox) key, $G$ the base point (unless stated otherwise). All DH calculations are modulo $p$. $\textrm{H}$ denotes a context-dependent hash function. $\textrm{Enc}_K(M)$ represents the symmetric encryption of $M$ under key $K$, and $\textrm{Enc}_K(IV,M,AD)$ further adds explicit IV and additional data for AEAD schemes. We also define $\mathrm{SignEC}_{P_\text{M}^-}(m_1G)$ as a shortcut to $\{m_1G,\mathrm{Sign}_{P_\text{M}^-}(\textrm{H}(r_{\text{C}},r_{\text{S}},m_1G))\}$.
Dashed arrows indicate a status, not message transfers. Double arrows represent messages encrypted under the established session.
\vspace{-.05in}

\tikzcdset{every label/.append style = {font = \scriptsize}}

\begin{figure}[h]
\begin{minipage}[t]{.38\textwidth}
\centering
\scriptsize
\begin{tikzcd}[column sep=2.4cm,row sep=10pt]
\txt{\textbf{C}} & \txt{\textbf{MB}} & \txt{\textbf{S}} \\
\ar[rr, "\txt{\TLSMsg{CH} ($r_{\text{C}}$)}"] & {} & {} \\
\ar[rr, leftarrow, "\txt{\TLSMsg{SH} ($r_{\text{S}}$)}"] & {} & {} \\
\ar[r, leftarrow, "\txt{\TLSMsg{CRT} ($P_\text{M}^+$)}"] & \ar[r, leftarrow, "\txt{\TLSMsg{CRT} ($P_\text{S}^+$)}"] & {} \\
\ar[r, leftarrow, "\txt{\TLSMsg{SKE}}", "\txt{$\Rdsh$ $\mathrm{SignEC}_{P_\text{M}^-}(m_1G)$}"'] & \ar[r, leftarrow, "\txt{\TLSMsg{SKE}}", "\txt{$\Rdsh$ $\mathrm{SignEC}_{P_\text{S}^-}(sG)$}"'] & {} \\[1em]
\ar[rr, leftarrow, "\txt{\TLSMsg{SHD}}"] & {} & {} \\
\ar[r, "\txt{\TLSMsg{CKE} ($cG$), \TLSMsg{CCS}}"] & \ar[r, "\txt{\TLSMsg{CKE} ($m_2G$), \TLSMsg{CCS}}"] & {} \\
\ar[r, "\txt{\TLSMsg{FIN} (client digest)}"] & \ar[r, "\txt{\TLSMsg{FIN} (middlebox digest)}"] & {} \\
\ar[r, leftarrow, "\txt{\TLSMsg{FIN} (middlebox digest)}"] & \ar[r, leftarrow, "\txt{\TLSMsg{FIN} (server digest)}"] & {} \\
\ar[r, dashed, leftrightarrow, "\txt{C-MB session}", "\txt{$\Rdsh$ Shared secret: $cm_1G$}"'] & \ar[r, dashed, leftrightarrow, "\txt{MB-S session}", "\txt{$\Rdsh$ Shared secret: $m_2sG$}"'] & {}
\end{tikzcd}
\vspace{-.1in}
\caption{TLS session splitting (TLS 1.2-ECDHE)}
\label{fig:tls-sess-split}
\end{minipage}
\begin{minipage}[t]{.30\textwidth}
\centering
\scriptsize
\begin{tikzcd}[column sep=1.6cm,row sep=9pt]
\txt{\textbf{C}} & \txt{\textbf{MB}} & \txt{\textbf{S}} \\
{} & \ar[r, dashed, leftrightarrow, "\txt{Mutually auth.\ session}"] & {} \\
\ar[rr, "\txt{\TLSMsg{CH}}"] & {} & {} \\
\ar[rr, leftarrow, "\txt{\TLSMsg{SH},\TLSMsg{CRT},\TLSMsg{SKE},\TLSMsg{SHD}}"] & {} & {} \\
\ar[rr, "\txt{\TLSMsg{CKE},\TLSMsg{CCS},\TLSMsg{FIN}}"] & {} & {} \\
\ar[rr, leftarrow, "\txt{\TLSMsg{CCS},\TLSMsg{FIN}}"] & {} & {} \\
\ar[rr, dashed, leftrightarrow, "\txt{C-S session}"] & {} & {} \\
\ar[rr, Rightarrow, "\txt{GET /img/1234 HTTP/1.1}"] & {} & {} \\
{} & \ar[r, Leftarrow, "\txt{\texttt{server\_write\_key}}"] & {} \\
{} & \ar[r, leftarrow, "\txt{ContentID=1234}"] & {} \\
{} & \ar[r, leftarrow, "\txt{tag1, tag2, ..., tagN}"] & {} \\
\ar[r, Leftarrow, "\txt{HTTP/1.1 200 OK}"] & {} & {}
\end{tikzcd}
\vspace{-.1in}
\caption{Barnraising~\cite{Lesniewski-LaasK05}}
\label{fig:barnraising}
\end{minipage}
\begin{minipage}[t]{.31\textwidth}
\centering
\scriptsize
\begin{tikzcd}[column sep=1.6cm,row sep=9pt]
\txt{\textbf{C}} & \txt{\textbf{MB}} & \txt{\textbf{S}} \\
{} & \ar[r, dashed, leftrightarrow, "\txt{Mutually auth.\ session}"] & {} \\
\ar[rr, "\txt{\TLSMsg{CH}}"] & {} & {} \\
\ar[rr, leftarrow, "\txt{\TLSMsg{SH},\TLSMsg{CRT},\TLSMsg{SKE},\TLSMsg{SHD}}"] & {} & {} \\
\ar[rr, "\txt{\TLSMsg{CKE},\TLSMsg{CCS},\TLSMsg{FIN}}"] & {} & {} \\
\ar[rr, leftarrow, "\txt{\TLSMsg{CCS},\TLSMsg{FIN}}"] & {} & {} \\
\ar[rr, dashed, leftrightarrow, "\txt{C-S session}"] & {} & {} \\
{} & \ar[r, Leftarrow, "\txt{\texttt{client\_write\_key}}"] & {} \\
\ar[rr, Rightarrow, "\txt{GET /img/1234 HTTP/1.1}"] & {} & {} \\
{} & \ar[r, Leftarrow, "\txt{\texttt{server\_write\_key}}"] & {} \\
{} & \ar[r, Leftarrow, "\txt{tag1, tag2, ..., tagN}"] & {} \\
\ar[r, Leftarrow, "\txt{HTTP/1.1 200 OK}"] & {} & {}
\end{tikzcd}
\vspace{-.18in}
\caption{Akamai's patent (Decryption+ One-Time Write Grant Upgrade)~\cite{gero2015splicing}}
\label{fig:akamai}
\end{minipage}
\end{figure}

\clearpage

\begin{figure}[h]
\centering
\begin{minipage}[t]{.95\textwidth}
\scriptsize
\begin{tikzcd}[column sep=3cm,row sep=9pt]
\txt{\textbf{C}} & \txt{\textbf{MB} ($P_\text{M}^-$)} & \txt{\textbf{S}} &[-2.6cm] \txt{\textbf{S} ($P_\text{S}^-$)} & \txt{\textbf{CA}} \\
{} & \ar[r, "\txt{Certificate Request ($P_\text{M}^+$)}"] & {} & \ar[r, "\txt{Certificate Request ($P_\text{S}^+$)}"] & {} \\
{} & \ar[r, leftarrow, "\txt{Certificate ($P_\text{M}^+$, CA: $0$)}", "\txt{$\Rdsh$ SAN=DNS.0:s1.domain.com}"'] & {} & \ar[r, leftarrow, "\txt{Certificate ($P_\text{S}^+$, CA: $1$)}", "\txt{$\Rdsh$ nameConstraints=\\permitted;DNS.0:domain.com}"'] & {} \\~\\
{} & \ar[r, leftarrow, "\txt{Certificate ($P_\text{M}^+$, CA: $0$)}", "\txt{$\Rdsh$ subj=/CN=domain.com/CN=proxy1,\\X509v3 Ext: ProxyCertInfo}"'] & {} & \ar[r, leftarrow, "\txt{Certificate ($P_\text{S}^+$, CA: $0$, DigSig)}", "\txt{$\Rdsh$ subj=/CN=domain.com}"'] & {} \\~\\
{} & \ar[r, leftarrow, "\txt{Certificate ($P_\text{M}^+$, CA: $0$)}", "\txt{$\Rdsh$ SAN=DNS.0:s1.domain.com}"'] & {} & \ar[r, leftarrow, "\txt{Certificate ($P_\text{S}^+$, CA: $0$)}", "\txt{$\Rdsh$ SAN=DNS.0:*.domain.com}"'] & {} \\[1em]
\ar[r, dashed, leftrightarrow, "\txt{TLS handshake w/ \TLSMsg{CRT} ($P_\text{M}^+$)}"] & {} & {} & {} & {}
\end{tikzcd}
\usetikzlibrary{decorations.pathreplacing}
\begin{tikzpicture}[remember picture,overlay]
\draw [dashed] (8,0.1) -- (8,4.2) node [black,midway,rotate=90,yshift=-.3cm] {\footnotesize $\downarrow$ Server cert.\ provisioning $\downarrow$};
\draw [decorate,decoration={brace,amplitude=10pt,mirror,raise=4pt},yshift=0pt]
(12.5,2.4) -- (12.5,3.25) node [right,xshift=.6cm,black,midway] {\footnotesize
RFC5280};
\draw [decorate,decoration={brace,amplitude=10pt,mirror,raise=4pt},yshift=0pt]
(12.5,1.4) -- (12.5,2.3) node [right,xshift=.6cm,black,midway] {\footnotesize
RFC3820};
\draw [decorate,decoration={brace,amplitude=10pt,mirror,raise=4pt},yshift=0pt]
(12.5,.7) -- (12.5,1.3) node[right,xshift=.6cm,black,midway] {\footnotesize
Chuat et al.};
\draw [decorate,decoration={brace,amplitude=10pt,raise=4pt},yshift=0pt]
(4,2.4) -- (4,3.25) node [left,xshift=-.6cm,black,midway] {\footnotesize
RFC5280};
\draw [decorate,decoration={brace,amplitude=10pt,raise=4pt},yshift=0pt]
(4,1.4) -- (4,2.3) node [left,xshift=-.6cm,black,midway] {\footnotesize
RFC3820};
\draw [decorate,decoration={brace,amplitude=10pt,raise=4pt},yshift=0pt]
(4,.7) -- (4,1.3) node [left,xshift=-.6cm,black,midway] {\footnotesize
Chuat et al.};
\end{tikzpicture}
\vspace{-.05in}
\caption{Name constraints (RFC5280~\cite{rfc5280}) and proxy certificates (RFC3820~\cite{rfc3820} and Chuat et al.~\cite{delegation-sok})}
\label{fig:proxy-cert}
\end{minipage}
\end{figure}

\begin{figure}[h]
\begin{minipage}[t]{.45\textwidth}
\centering
\scriptsize
\begin{tikzcd}[column sep=2.2cm,row sep=10pt]
\txt{\textbf{C}} & \txt{\textbf{MB} ($m$)} & \txt{\textbf{S}} \\
{} & \ar[r, dashed, leftrightarrow, "\txt{Mutually auth.\ session}"] & {} \\
\ar[r, "\txt{\TLSMsg{CH} ($r_{\text{C}}$)}"] & \ar[r, "\txt{$\textrm{H}(r_{\text{C}},r_{\text{S}},mG)$}"] & {} \\
\ar[r, leftarrow, "\txt{\TLSMsg{SH} ($r_{\text{S}}$), \TLSMsg{CRT} ($P_\text{S}^+$)}"] & {} & {} \\
\ar[r, leftarrow, "\txt{\TLSMsg{SKE},\TLSMsg{SHD}}"] & \ar[r, leftarrow, "\txt{$\textrm{Sign}_{P_\text{S}^-}(\textrm{H}(r_{\text{C}},r_{\text{S}},mG))$}"] & {} \\
\ar[r, "\txt{\TLSMsg{CKE},\TLSMsg{CCS},\TLSMsg{FIN}}"] & {} & {} \\
\ar[r, leftarrow, "\txt{\TLSMsg{CCS},\TLSMsg{FIN}}"] & {} & {} \\
\ar[r, dashed, leftrightarrow, "\txt{C-MB session}"] & {} & {}
\end{tikzcd}
\vspace{-.1in}
\caption{Keyless SSL (TLS 1.2-ECDHE)~\cite{keyless-ssl}}
\label{fig:keyless-ssl}
\end{minipage}
\begin{minipage}[t]{.5\textwidth}
\centering
\scriptsize
\begin{tikzcd}[column sep=2.6cm,row sep=10pt]
\txt{\textbf{C}} & \txt{\textbf{MB} ($m, S$)} & \txt{\textbf{S}} \\
{} & \ar[r, dashed, leftrightarrow, "\txt{EAP-TLS exported key $k$}"] & {} \\
\ar[r, "\txt{\TLSMsg{CH} ($r_{\text{C}}$)}"] & \ar[r, "\txt{$r_{\text{C}},S,mG$}"] & {} \\
\ar[r, leftarrow, "\txt{\TLSMsg{SH} ($\varphi_k(S)$), \TLSMsg{CRT} ($P_\text{S}^+$)}"] & {} & {} \\
\ar[r, leftarrow, "\txt{\TLSMsg{SKE},\TLSMsg{SHD}}"] & \ar[r, leftarrow, "\txt{$\textrm{Sign}_{P_\text{S}^-}(\textrm{H}(r_{\text{C}},\varphi_k(S),mG))$}"] & {} \\
\ar[r, "\txt{\TLSMsg{CKE},\TLSMsg{CCS},\TLSMsg{FIN}}"] & {} & {} \\
\ar[r, leftarrow, "\txt{\TLSMsg{CCS},\TLSMsg{FIN}}"] & {} & {} \\
\ar[r, dashed, leftrightarrow, "\txt{C-MB session}"] & {} & {}
\end{tikzcd}
\vspace{-.1in}
\caption{LURK (TLS 1.2-ECDHE)~\cite{draft-mglt-lurk-tls12-05}}
\label{fig:lurk}
\end{minipage}
\end{figure}

\begin{figure}[h]
\begin{minipage}{.45\textwidth}
\centering
\scriptsize
\begin{tikzcd}[column sep=2.5cm,row sep=12pt]
\txt{\textbf{C}} & \txt{\textbf{MB} ($m$)} & \txt{\textbf{S} ($DC^-$)} \\[-.5em]
{} & \ar[r, dashed, leftrightarrow, "\txt{Mutually auth.\ session}"] & {} \\
{} & \ar[r, Leftarrow, "\txt{$DC^-, \textrm{Sign}_{P_\text{S}^-}(DC^+)$}"] & {} \\
\ar[r, "\txt{\TLSMsg{CH} ($r_{\text{C}}, cG$)}"] & {} & {} \\
\ar[r, leftarrow, "\txt{\TLSMsg{SH} ($r_{\text{S}}, mG$), \TLSMsg{EE}}"] & {} & {} \\
\ar[r, leftarrow, "\txt{\TLSMsg{CRT} ($P_\text{S}^+, \textrm{delegated\_cred}$:}", "\txt{$\Rdsh \{DC^+, \textrm{Sign}_{P_\text{S}^-}(DC^+)\}$)}"'] & {} & {} \\[1.5em]
\ar[r, leftarrow, "\txt{\TLSMsg{CV} ($\textrm{Sign}_{DC^-}(\tau)$)}"] & {} & {} \\
\ar[r, leftarrow, "\txt{\TLSMsg{FIN}}"] & {} & {} \\
\ar[r, "\txt{\TLSMsg{FIN}}"] & {} & {} \\
\ar[r, dashed, leftrightarrow, "\txt{C-MB session}"] & {} & {}
\end{tikzcd}
\vspace{-.05in}
\caption{Delegated Credentials (TLS 1.3)~\cite{draft-ietf-tls-subcerts-09}}
\label{fig:delegated-credentials}
\end{minipage}\hspace{-35mm}
\begin{minipage}{.3\textwidth}
\centering
\scriptsize
\vspace{11mm}
\begin{tikzcd}[column sep=1cm,row sep=10pt]
\txt{\textbf{C} ($c$)} & \txt{\textbf{MB} ($d_1$)} & \txt{\textbf{S} ($s,d_1G,d_2$)} \\
\ar[rr, "\txt{\TLSMsg{CH} ($r_{\text{C}},cG,\textrm{tls\_visibility}$:\{\})}"] & {} & {} \\
\ar[rr, leftarrow, "\txt{\TLSMsg{SH} ($r_{\text{S}},sG,\textrm{tls\_visibility}$:}", "\txt{$\Rdsh$ \{$d_2G,\textrm{Enc}_{Ke}(\text{\sffamily{handshake\_secret}})$\})}"'] & {} & {} \\[1.5em]
\ar[rr, leftarrow, "\txt{\TLSMsg{EE},\TLSMsg{CRT} ($P_\text{S}^+$),\TLSMsg{CV},\TLSMsg{FIN}}"] & {} & {} \\
\ar[rr, "\txt{\TLSMsg{FIN}}"] & {} & {} \\
\ar[rr, dashed, leftrightarrow, "\txt{C-S session}"] & {} & {}
\end{tikzcd}
\begin{tikzpicture}[remember picture,overlay]
\draw [dashed] (-2.5,-0.5) -- (-2.5,3.7);
\draw [dashed] (-2.5,3.7) -- (2.8,3.7);
\end{tikzpicture}
\vspace{-.1in}
\caption{\textsf{tls\_visibility} (TLS 1.3)~\cite{draft-rhrd-tls-tls13-visibility-01}}
\label{fig:tls-visibility}
\end{minipage}
\begin{minipage}{.45\textwidth}
\centering
\scriptsize
\begin{tikzcd}[column sep=2.5cm,row sep=10pt]
\txt{\textbf{C}} & \txt{\textbf{MB}} & \txt{\textbf{S}} \\
\ar[rr, "\txt{\TLSMsg{CH} ($r_{\text{C}},\textrm{efgh:\{}P_\text{M}^+\textrm{, policy\}}$)}"] & {} & {} \\
\ar[rr, leftarrow, "\txt{\TLSMsg{SH} ($r_{\text{S}},\textrm{efgh:\{}P_\text{M}^+\textrm{, policy\}}$)}"] & {} & {} \\
\ar[rr, leftarrow, "\txt{\TLSMsg{CRT} ($P_\text{S}^+$),\TLSMsg{SKE},\TLSMsg{SHD}}"] & {} & {} \\
\ar[rr, "\txt{\TLSMsg{CKE},\TLSMsg{CCS},\TLSMsg{FIN}}"] & {} & {} \\
\ar[rr, leftarrow, "\txt{\TLSMsg{FIN}}"] & {} & {} \\
\ar[rr, dashed, leftrightarrow, "\txt{C-S session, exported keys $G,K_{CS}$}"] & {} & {} \\
\ar[r, leftrightarrow, "\txt{DH key exchange: $K_{CP}$}"] & \ar[r, leftrightarrow, "\txt{DH key exchange: $K_{PS}$}"] & {} \\
\ar[r, "\txt{$\textrm{Enc}_{K_{CP}}(G)$}"] & {} & {} \\
\ar[rr, "\txt{$n_1,n_2,H\textrm{:false},C_G\textrm{:Enc}_{G}(n_1,\textrm{``http''},H)$,}", "\txt{$\Rdsh C_{K_{CS}}\textrm{:Enc}_{K_{CS}}(n_2,\textrm{``GET /img?secret=}\dots\textrm{''},C_G)$}"'] & {} & {} \\[1em]
\ar[rr, leftarrow, "\txt{$n_3,n_4,H\textrm{:true},C_G\textrm{:Enc}_{G}(n_3,\textrm{``HTTP/1.1 200 OK''},H)$,}", "\txt{$\Rdsh C_{K_{CS}}\textrm{:Enc}_{K_{CS}}(n_4,\phi,C_G)$}"'] & {} & {} \\
\end{tikzcd}
\vspace{-.05in}
\caption{EFGH~\cite{efgh}}
\label{fig:efgh}
\end{minipage}
\end{figure}

\clearpage

\begin{figure}
\begin{minipage}{.56\textwidth}
\centering
\scriptsize
\begin{tikzcd}[column sep=3cm,row sep=10pt]
\txt{\textbf{C} ($c$)} & \txt{\textbf{MB} ($m_1,m_2$)} & \txt{\textbf{S} ($s$)} \\
\ar[rr, "\txt{\TLSMsg{CH} ($r_{\text{C}}$, MiddleboxList:\{ctx:\{``request'',``response''\},p:10.0.0.2:4433\})}"] & {} & {} \\[.5em]
\ar[rr, leftarrow, "\txt{\TLSMsg{SH} ($r_{\text{S}}$),\TLSMsg{CRT} ($P_\text{S}^+$),\TLSMsg{SKE} ($\mathrm{SignEC}_{P_\text{S}^-}(sG)$),\TLSMsg{SHD}}"] & {} & {} \\
\ar[r, leftarrow, "\txt{\TLSMsg{SH} ($r_{\text{M}}$),\TLSMsg{CRT} ($P_\text{M}^+$)}"] & \ar[r, "\txt{\TLSMsg{SH} ($r_{\text{M}}$),\TLSMsg{CRT} ($P_\text{M}^+$)}"] & {} \\[.5em]
\ar[r, leftarrow, "\txt{\TLSMsg{SKE} ($\mathrm{SignEC}_{P_\text{M}^-}(m_1G)$),\TLSMsg{SHD}}"] & \ar[r, "\txt{\TLSMsg{SKE} ($\mathrm{SignEC}_{P_\text{M}^-}(m_2G)$),\TLSMsg{SHD}}"] & {} \\
\ar[rr, "\txt{\TLSMsg{CKE} ($cG$),\TLSMsg{CCS},\TLSMsg{FIN}}"] & {} & {} \\
\ar[rr, "\txt{\TLSMsg{MKM (MB)} (ctx0:$\phi$,ctx1:$\textrm{Enc}_{K1_{\text{C,MB}}}(\{K1_\text{rd}^{\text{C}},K1_\text{wr}^{\text{C}}\})$}"] & {} & {} \\
\ar[rr, "\txt{\TLSMsg{MKM (S)} (ctx0:$\textrm{Enc}_{K_{\text{C,S}}}(\{K0_\text{rd}^{\text{C}},K0_\text{wr}^{\text{C}}\})$,ctx1:$\textrm{Enc}_{K_{\text{C,S}}}(\{K1_\text{rd}^{\text{C}},K1_\text{wr}^{\text{C}}\})$}"] & {} & {} \\
\ar[rr, "\txt{\TLSMsg{CCS},\TLSMsg{FIN}}"] & {} & {} \\
\ar[rr, leftarrow, "\txt{\TLSMsg{MKM (MB)} (ctx0:$\phi$,ctx1:$\textrm{Enc}_{K_{\text{S,MB}}}(\{K1_\text{rd}^{\text{S}},K1_\text{wr}^{\text{S}}\})$}"] & {} & {} \\
\ar[rr, leftarrow, "\txt{\TLSMsg{MKM (C)} (ctx0:$\textrm{Enc}_{K_{\text{C,S}}}(\{K0_\text{rd}^{\text{S}},K0_\text{wr}^{\text{S}}\})$,ctx1:$\textrm{Enc}_{K_{\text{C,S}}}(\{K1_\text{rd}^{\text{S}},K1_\text{wr}^{\text{S}}\})$}"] & {} & {} \\
\ar[rr, leftarrow, "\txt{\TLSMsg{CCS},\TLSMsg{FIN}}"] & {} & {} \\
\ar[rr, dashed, leftrightarrow, "\txt{C-M-S session, context key sets: $K_{\text{C,S}},K0_{\text{rd}},K0_{\text{wr}},K1_{\text{rd}},K1_{\text{wr}}$}"] & {} & {} \\
\ar[rr, "\txt{$E$:$\textrm{Enc}_{K0_{\text{rd}}}(\textrm{``GET /img?secret=}\dots\textrm{''}),\textrm{MAC}_{K_{\text{C,S}}}(E),\textrm{MAC}_{K0_{\text{rd}}}(E),\textrm{MAC}_{K0_{\text{wr}}}(E)$}"] & {} & {} \\
\ar[rr, leftarrow, "\txt{$E$:$\textrm{Enc}_{K1_{\text{rd}}}(\textrm{``HTTP/1.1 200 OK}\dots\textrm{''}),\textrm{MAC}_{K_{\text{C,S}}}(E),\textrm{MAC}_{K1_{\text{rd}}}(E),\textrm{MAC}_{K1_{\text{wr}}}(E)$}"] & {} & {}
\end{tikzcd}
\vspace{-.1in}
\caption{mcTLS (1 MB, ctx0: no MB access, ctx1: R/W access)~\cite{mctls}}
\label{fig:mctls}
\vspace{.2in}
\begin{tikzcd}[column sep=3.2cm,row sep=10pt]
\txt{\textbf{C} ($c$)} & \txt{\textbf{MB} ($m$)} & \txt{\textbf{S} ($s,s'$)} \\
\ar[rr, "\txt{\TLSMsg{CH} ($r_{\text{C}}$, MiddleboxSupportExtension)}"] & {} & {} \\
{} & \ar[r, "\txt{\TLSMsg{MBTLSMiddleboxAnnouncement}}"] & {} \\[.5em]
\ar[rr, leftarrow, "\txt{\TLSMsg{SH} ($r_{\text{S}}$, MiddleboxSupportExtension),\TLSMsg{CRT},\TLSMsg{SKE} ($\mathrm{SignEC}_{P_\text{S}^-}(sG)$),\TLSMsg{SHD}}"] & {} & {} \\
{} & \ar[r, leftarrow, "\txt{\TLSMsg{CH} ($r_{\text{S}'}$)}"] & {} \\
\ar[rr, "\txt{\TLSMsg{CKE} ($cG$),\TLSMsg{CCS},\TLSMsg{FIN}}"] & {} & {} \\
{} & \ar[r, "\txt{\TLSMsg{SH} ($r_{\text{M}}$),\TLSMsg{CRT} ($P_\text{M}^+$),}", "\txt{$\Rdsh$ \TLSMsg{SKE} ($\mathrm{SignEC}_{P_\text{M}^-}(mG)$),\TLSMsg{SHD}}"'] & {} & {} \\[1.5em]
{} & \ar[r, "\txt{\TLSMsg{SGX Attestation}}"] & {} \\
\ar[rr, leftarrow, "\txt{\TLSMsg{CCS},\TLSMsg{FIN}}"] & {} & {} \\
\ar[rr, dashed, leftrightarrow, "\txt{C-S session (shared key $K_{\text{C,S}}$)}"] & {} & {} \\
{} & \ar[r, leftarrow, "\txt{\TLSMsg{CKE} ($s'G$),\TLSMsg{CCS},\TLSMsg{FIN}}"] & {} & {} \\
{} & \ar[r, leftarrow, "\txt{\TLSMsg{MBTLSKeyMaterial} ($K_{\text{C,S}}$)}"] & {} \\
{} & \ar[r, leftarrow, "\txt{\TLSMsg{MBTLSKeyMaterial} ($K_{\text{MB,S}}$)}"] & {} \\
{} & \ar[r, "\txt{\TLSMsg{CCS},\TLSMsg{FIN}}"] & {} & {} \\
\ar[r, "\txt{$\textrm{Enc}_{K_{\text{C,S}}}(\textrm{``GET /img/1234 HTTP/1.1''})$}"] & \ar[r, "\txt{$\textrm{Enc}_{K_{\text{MB,S}}}(\textrm{``GET /img/1234 HTTP/1.1''})$}"] & {} \\
\ar[r, leftarrow, "\txt{$\textrm{Enc}_{K_{\text{C,S}}}(\textrm{``HTTP/1.1 200 OK}\dots\textrm{''})$}"] & \ar[r, leftarrow, "\txt{$\textrm{Enc}_{K_{\text{MB,S}}}(\textrm{``HTTP/1.1 200 OK}\dots\textrm{''})$}"] & {}
\end{tikzcd}
\vspace{-.1in}
\caption{mbTLS (1 server-side MB)~\cite{mbtls}}
\label{fig:mbtls}
\end{minipage}
\begin{minipage}{.4\linewidth}
\scriptsize
\[
\rotatebox{90}{%
\begin{tikzcd}[column sep=6cm,row sep=10pt,ampersand replacement=\&]
\txt{\textbf{C} ($c',c$)} \& \txt{\textbf{MB} ($m',m_1,m_2$)} \& \txt{\textbf{S} ($s',s$)} \\
\ar[r, "\txt{\TLSMsg{CH} ($r_{\text{C}}$, MbA:\{$c'G'$\})}"] \& \ar[r, "\txt{\TLSMsg{CH} ($r_{\text{M}}$, MbA:\{$c'G',m'G'$\})}"] \& {} \\
\ar[r, leftarrow, "\txt{\TLSMsg{SH} ($r_{\text{S}}$, MbA:\{$s'G',m'G'$\}),\TLSMsg{CRT} ($P_\text{S}^+,P_\text{M}^+$)}"] \& \ar[r, leftarrow, "\txt{\TLSMsg{SH} ($r_{\text{S}}$, MbA:\{$s'G'$\}),\TLSMsg{CRT} ($P_\text{S}^+$)}"] \& {} \\[0.5em]
\ar[r, leftarrow, "\txt{\TLSMsg{SKE} ($\mathrm{SignEC}_{P_\text{M}^-}(m_2G)$),\TLSMsg{SHD}}"] \& \ar[r, leftarrow, "\txt{\TLSMsg{SKE} ($\mathrm{SignEC}_{P_\text{S}^-}(sG)$),\TLSMsg{SHD}}"] \& {} \\
\ar[r, "\txt{\TLSMsg{CKE} ($cG$),\TLSMsg{CCS},\TLSMsg{FIN}}"] \& \ar[r, "\txt{\TLSMsg{CKE} ($m_1G$),\TLSMsg{CCS},\TLSMsg{FIN}}"] \& {} \\
\ar[r, leftarrow, "\txt{\TLSMsg{CCS},\TLSMsg{FIN}}"] \& \ar[r, leftarrow, "\txt{\TLSMsg{CCS},\TLSMsg{FIN}}"] \& {} \\
\ar[r, dashed, leftrightarrow, "\txt{C-MB session, $ak_{\text{MB,C}}=\textrm{PRF}(c'm'G',\textrm{label},s'G'\|c'G')$}"] \& \ar[r, dashed, leftrightarrow, "\txt{MB-S session, $ak_{\text{MB,S}}=\textrm{PRF}(m's'G',\textrm{label},s'G'\|c'G')$}"] \& {} \\[0.5em]
\ar[r, leftarrow, "\txt{\TLSMsg{ExtFin} ($SPB_{\text{S}},SPB_{\text{MB}}\textrm{:}\{\mathrm{H}(ak_{\text{MB,C}}),p_{\text{MB,S}}, \mathrm{Sign}_{P_\text{M}^-}(\mathrm{MAC}_{ak_{\text{MB,C}}}($}", "\txt{$\Rdsh$ $\{\mathrm{0303},\mathrm{c02c},\mathrm{H}(\texttt{master\_secret}),\texttt{verify\_data}\}_{\text{C}\rightarrow\text{MB}}\|p_{\text{MB,S}}))\})$}"'] \& \ar[r, leftarrow, "\txt{\TLSMsg{ExtFin} ($SPB_{\text{S}}\textrm{:}\{\mathrm{H}(ak_{\text{S,C}}), \mathrm{Sign}_{P_\text{S}^-}(\mathrm{MAC}_{ak_{\text{S,C}}}(p_{\text{MB,S}}$:}", "\txt{$\Rdsh$ $\{\mathrm{0303},\mathrm{c02f},\mathrm{H}(\texttt{master\_secret}),\texttt{verify\_data}\}_{\text{MB}\rightarrow\text{S}}))\})$}"'] \& {} \\[1.5em]
\ar[r, Rightarrow, "\txt{$q$:$\textrm{``GET /img/1234 HTTP/1.1''},\mathrm{MAC}_{ak_{\text{S,C}}}(\mathrm{H}(q))$}"] \& \ar[r, Rightarrow, "\txt{$q,\mathrm{MAC}_{ak_{\text{MB,S}}}(\mathrm{MAC}_{ak_{\text{S,C}}}(\mathrm{H}(q)))$}"] \& {} \\
\ar[r, Leftarrow, "\txt{$r,\mathrm{MAC}_{ak_{\text{MB,C}}}(\mathrm{MAC}_{ak_{\text{S,C}}}(\mathrm{H}(r)))$}"] \& \ar[r, Leftarrow, "\txt{$r$:$\textrm{``HTTP/1.1 200 OK}\dots\textrm{''},\mathrm{MAC}_{ak_{\text{S,C}}}(\mathrm{H}(r))$}"] \& {}
\end{tikzcd}
}
\]
\vspace{-.15in}
\caption{maTLS (1 MB, no mod.)~\cite{matls} (1 MB)}
\label{fig:matls}
\end{minipage}
\end{figure}
\end{document}